%% file: main.tex
\documentclass[fleqn,usenatbib]{mnras}

\usepackage[T1]{fontenc}
\usepackage{ae,aecompl}

\usepackage{graphicx}	
\usepackage{amsmath}	
\usepackage{amssymb}	
\usepackage{gensymb}
\usepackage{mathrsfs}
\usepackage{txfonts, mathptmx}
\usepackage{dblfloatfix}
\usepackage{bm}
\usepackage{footmisc}
\usepackage{relsize}
\usepackage{booktabs}
\usepackage{hyperref}
\usepackage{units}
\usepackage{subfiles}

\newcommand{\kms}{\, \mathrm{km} \, \mathrm{s}^{-1}}
\newcommand{\feh}{\mathrm{[Fe/H]}}
\newcommand{\fehavr}{\overline{ \feh }}
\newcommand{\ra}[1]{\renewcommand{\arraystretch}{#1}}

\title[Chemodynamic Populations in Ursa Minor]{Multiple Chemodynamic Stellar Populations of the Ursa Minor Dwarf Spheroidal Galaxy\thanks{The data presented herein were obtained at the W.M. Keck Observatory, which is operated as a scientific partnership among the California Institute of Technology, the University of California and the National Aeronautics and Space Administration. The Observatory was made possible by the generous financial support of the W.M. Keck Foundation.}}

\author[A. B. Pace et al.]{
\parbox{\textwidth}{
\Large
Andrew B. Pace$^{1,2, 3}$ \thanks{E-mail: apace@andrew.cmu.edu },
Manoj Kaplinghat$^3$,
Evan Kirby$^{4}$,
Joshua D. Simon$^{5}$,
Erik Tollerud$^{6}$,
Ricardo R. Mu{\~n}oz$^{7}$,
Patrick C{\^o}t{\'e}$^{8}$,
S. G. Djorgovski$^{4}$,
Marla Geha$^{9}$
 \\ 
\small {\it 
$^{1}$McWilliams Center for Cosmology, Carnegie Mellon University, 5000 Forbes Ave, Pittsburgh, PA 15213, USA \\
$^2$George P. and Cynthia Woods Mitchell Institute for Fundamental Physics and Astronomy, and 
Department of Physics and Astronomy, Texas A\&M University, College Station, TX 77843, USA \\
$^3$Center for Cosmology, Department of Physics and Astronomy, University of California, Irvine, CA 92697, USA \\
$^{4}$California Institute of Technology, 1200 E. California Blvd., MC 249-17, Pasadena, CA 91125, USA \\
$^{5}$Observatories of the Carnegie Institution for Science, 813 Santa Barbara Street, Pasadena, CA 91101, USA\\
$^{6}$Space Telescope Science Institute, 3700 San Martin Drive, Baltimore, MD 21218, USA\\
$^{7}$Departamento de Astronom{\'i}a, Universidad de Chile, Camino del Observatorio 1515, Las Condes, Santiago, Chile \\
$^{8}$National Research Council of Canada, Herzberg Astronomy \& Astrophysics Research Centre, 5071 W. Saanich Road, Victoria, BC, V9E 2E7, Canada \\
$^{9}$ Astronomy Department, Yale University, New Haven, CT 06520, USA\\
}
}
}

\date{Accepted XXX. Received YYY; in original form ZZZ}

\pubyear{2020}

\begin{document}
\label{firstpage}
\pagerange{\pageref{firstpage}--\pageref{lastpage}}
\maketitle

\begin{abstract}
We present a Bayesian method to identify multiple (chemodynamic) stellar populations in dwarf spheroidal galaxies (dSphs) using velocity, metallicity, and positional stellar data without the assumption of spherical symmetry. 
We apply this method to a new Keck/DEIMOS spectroscopic survey of the Ursa Minor (UMi) dSph.  
We identify 892 likely members, making this the largest UMi sample with line-of-sight velocity and metallicity measurements.
Our Bayesian method detects two distinct chemodynamic populations with high significance (in logarithmic Bayes' factor, $\ln{B}\sim33$). 
The metal-rich ($[{\rm Fe/H}]=-2.05\pm0.03$) population is kinematically colder  (radial velocity dispersion of $\sigma_v=4.9_{-1.0}^{+0.8} \kms$) and more centrally concentrated than the metal-poor ($[{\rm Fe/H}]=-2.29_{-0.06}^{+0.05}$) and kinematically hotter population ($\sigma_v =11.5_{-0.8}^{+0.9}\kms$).  
Furthermore, we apply the same analysis to an independent MMT/Hectochelle data set and confirm the existence of two chemodynamic populations in UMi. 
In both data sets, the metal-rich population is significantly flattened ($\epsilon=0.75\pm0.03$) and the metal-poor population is closer to spherical ($\epsilon=0.33_{-0.09}^{+0.12}$).
Despite the presence of two populations, we are unable to robustly estimate the slope of the dynamical mass profile. 
We found hints for prolate rotation of order $\sim 2 \kms$ in the MMT data set, but further observations are required to verify this. 
The flattened metal-rich population invalidates assumptions built into simple dynamical mass estimators, so we computed new astrophysical dark matter annihilation (J) and decay profiles based on the rounder, hotter metal-poor population and inferred $\log_{10}{(J(0.5\degree)/{\rm GeV^{2} \, cm^{-5}})}\approx19.1$ for the Keck data set. Our results paint a more complex picture of the evolution of Ursa Minor than previously discussed.
\end{abstract}


\section{Introduction}

The distribution of dark matter within galaxies is a key test for the $\Lambda$CDM (cosmological constant + cold dark matter) cosmological model.
Dark matter-only simulations predict that dark matter halos have cuspy inner density slopes that scale as $\rho_{\rm DM} \sim r^{-1}$ at small radii \citep{Navarro1996ApJ...462..563N, Navarro1997ApJ...490..493N}.
Observations of dwarf, spiral, and low surface brightness galaxies infer  shallower profiles  \citep[e.g.,][]{deBlock2002, Simon2005, KDN_2006, Kuzio_de_Naray2008, deblok2008, Oh2011, Adams2014, Oh2015AJ....149..180O, Relatores2019ApJ...887...94R}.  
Solutions to this disagreement generally fall into two categories: the inclusion of baryonic effects \citep[e.g.,][]{Navarro1996MNRAS.283L..72N, Governato2010, Pontzen2012, Governato2012, Penarrubia2012, Brooks2013, Onorbe2015} or a dark matter model differing from the canonical cold and non-interacting model \citep[e.g., recent work by][]{Rocha2013,Peter2013MNRAS.430..105P, Lovell2014, Horiuchi2014, Kaplinghat2014, Wang2014, Kaplinghat2016PhRvL.116d1302K,Abazajian2014, Kamada2017PhRvL.119k1102K}.  
There is ongoing debate in the literature over the validity of both solutions.
The Milky Way (MW) dwarf spheroidal   galaxies (dSph) have low stellar masses and are highly dark matter dominated systems \citep{McConnachie2012AJ....144....4M, Simon2019ARA&A..57..375S}.  Accordingly they are excellent laboratories to distinguish between these solutions. 

The MW dSphs are close enough for photometric and spectroscopic analysis of individual stars.
Analysis of color-magnitude diagrams has revealed that the brighter ``classical'' dSphs ($L_{\rm V} \gtrsim 10^5 M_{\odot}$)  have extended star formation histories \citep{Weisz2014ApJ...789..147W}.  
The  spatial distributions of different stellar populations 
in dSphs may vary as a function of age or metallicity, depending on the dynamical history of the galaxy and the evolution of its gas distribution over time. 
For example, in similar galaxies red horizontal branch (younger and more metal-rich) stars are generally more centrally concentrated than the older blue horizontal branch stars \citep{Harbeck2001}.  
Large spectroscopic surveys with accurate velocity and metallicity measurements have shown that the stellar kinematics are distinct  between the metal-poor and metal-rich populations   \citep[e.g.,][]{Tolstoy2004ApJ...617L.119T, Battaglia2006A&A...459..423B, McConnachie2007MNRAS.380L..75M, Battaglia2008ApJ...681L..13B, Walker2011ApJ...742...20W, Hendricks2014A&A...572A..82H, Kordopatis2016MNRAS.457.1299K}.

The MW dSphs are dispersion supported, gas-free systems and among the closest objects for which the motions of individual stars can be utilized for dynamical analysis. 
Unfortunately, a direct inference of the inner mass slope  is hampered by the degeneracy between the  mass profile and stellar anisotropy. 
One promising approach to breaking this degeneracy in the dSphs is to utilize the dynamics of multiple chemodynamic stellar populations \citep{Battaglia2008ApJ...681L..13B, Walker2011ApJ...742...20W, Read2017MNRAS.471.4541R}.

Thus far, the dynamics of multiple stellar populations have been utilized in three dSphs: Fornax \citep{Walker2011ApJ...742...20W, Amorisco2013MNRAS.429L..89A}, Sculptor \citep{Battaglia2008ApJ...681L..13B, Walker2011ApJ...742...20W, Amorisco2012ApJ...756L...2A, Agnello2012ApJ...754L..39A, Strigari2017ApJ...838..123S, Zhu2016MNRAS.463.1117Z}, and Carina \citep{Hayashi2018MNRAS.481..250H} to infer the mass slope of the dark matter halo.
Most results favor a `cored' halo over `cuspy' halo, but whether the `cuspy' solution is excluded \citep{Walker2011ApJ...742...20W, Amorisco2012ApJ...756L...2A, Agnello2012ApJ...754L..39A}, just disfavored  \citep{Battaglia2008ApJ...681L..13B, Amorisco2013MNRAS.429L..89A} or consistent \citep{Strigari2017ApJ...838..123S} is under debate.  
Finding additional dSphs with multiple chemodynamic populations will assist in determining whether the inner dark matter profile in dSphs is a `cusp' or a `core.'

In this paper, we present results from a Keck/DEIMOS spectroscopic survey of the Ursa Minor (UMi) dSph that shows, with high significance, two chemodynamic stellar populations.  
In Section~\ref{section:data}, we discuss the observations, data reduction, velocity and metallicity measurements, color-magnitude selection, and final catalog selection. 
In Section~\ref{section:method}, we present our statistical methodology, likelihood, and methods to separate foreground Milky Way stars and separate and identify multiple stellar populations.
In Section~\ref{section:results}, we present our main results on  the properties of the chemodynamic populations of UMi, verify our chemodynamic results with an independent MMT/Hectochelle data set \citep{Spencer2018AJ....156..257S}, and search for stellar rotation.
In Section~\ref{section:slope}, we discuss the inner slope of the mass profile. In Section~\ref{section:chemo_compare} compare our results in UMi to other dSphs  and in Section~\ref{section:j_factor} compute the astrophysical components  for studies of the indirect detection of dark matter.  
In Section~\ref{sec:conclusion}, we conclude and summarize our results.

\section{Data}
\label{section:data}

\subsection{Observations and Target Selection}
\label{section:obs}

\begin{table*}
\begin{center}
\label{table:mask}
\caption{Observation Log}
\begin{tabular}{cccccc}
\hline
Slitmask & No. of Targets & Date & Airmass & Seeing & Exposures \\
\hline
uss-1 & 68 & 2012 April 19 &  1.58 & 1.1\arcsec & 3$\times$1020 s\\
      & 68& 2012 April 23 &  1.60&  0.8\arcsec & 1$\times$1020 s\\
uss-2 & 57 & 2012 April 19  & 1.74 & 1.0\arcsec & 2$\times$1020 s, 600 s\\
uss-3 & 74 & 2012 April 21 &  1.55 & 0.5\arcsec & 3$\times$960 s\\
uss-4& 66 &  2012 April 21 &  1.70 & 0.7\arcsec  &3$\times$960 s, 480 s\\
uss-5 &27& 2012 April 21 &  1.49 & 0.5\arcsec & 2$\times$960 s\\
uss-6 & 13 & 2012 April 22  & 1.49 & 0.7\arcsec &2$\times$960 s, 900 s\\
uss-7 &17& 2012 April 23  & 1.49  &0.9\arcsec & 2$\times$1020 s\\
uss-8 & 57 & 2012 April 22  & 1.56 & 0.9\arcsec  &2$\times$1080 s, 1170 s\\
uss-9 &24& 2012 April 23 &  1.55 & 0.7\arcsec & 1$\times$1080 s, 1020 s\\
uss-10 & 65 &2012 April 22 &  1.47&  0.8\arcsec&  3$\times$1020 s\\
uss-11& 56 & 2012 April 21&  1.48&  0.5\arcsec & 3$\times$960 s\\
uss-12& 54 & 2012 April 23 &  1.47 & 0.9\arcsec &  3$\times$1020 s\\ 
\hline
\end{tabular}
\par
\end{center}
\end{table*}

Spectroscopic observations were carried out February 22--23 2009
\citep[first presented by][]{Kirby2010ApJS..191..352K}, May 11--12 2010 \citep[first presented by][]{Kirby2018ApJS..237...18K}, and April 20--22
2012 (not previously published) on the Keck/DEIMOS spectrograph \citep{DEIMOS2003}.  
All these observations used the
1200G diffraction grating, which has a groove spacing of
1200~mm$^{-1}$ and a blaze wavelength of 7760~\AA\@.  The grating was
tilted such that the typical central wavelength of a spectrum was
7800~\AA, and the typical wavelength range was about 2600~\AA\@.  In
practice, the wavelength range for each spectrum varied by up to
300~\AA\ depending on the location of the slit along the dispersion
axis.  The grating was used in first order, and higher orders were
blocked with the OG550 filter.  DEIMOS has a flexure compensation
system that keeps the wavelength calibration stable to within $\sim
0.1$~\AA\ over a full night.
The spectral resolution is $\Delta(\lambda) \sim 1.2$~\AA, which translates to $R \sim 7000$ at the Ca triplet around 8500~\AA\@.

Spectroscopic targets were selected from various photometric catalogs.
Where the slitmask design constraints forced a choice among multiple
candidates, we prioritized stars on the red giant branch (RGB)\@.
\citet{Kirby2010ApJS..191..352K} described the slitmask design for the 2009
observations.  Targets were selected from \citeauthor{Bellazzini2002AJ....124.3222B}'s
(\citeyear{Bellazzini2002AJ....124.3222B}) photometric catalog in the $V$ and $I$
filters.  We used the color--magnitude diagram (CMD) in conjunction
with Yonsei-Yale isochrones \citep{Demarque2004} to inform the
selection.  Targets were selected between a blue bound and red bound.
The blue bound was 0.1~mag bluer in de-reddened $(V-I)_0$ than a 2~Gyr
isochrone with ${\rm [Fe/H]} = -3.76$ and ${\rm [\alpha/Fe]} = 0.0$.
The red bound was a 14~Gyr isochrones with ${\rm [Fe/H]} = +0.05$ and
${\rm [\alpha/Fe]} = +0.3$.  Horizontal branch (HB) stars were also
selected from a box in the CMD: $20.5 > I_0 > 19.0$ and $-0.20 <
(V-I)_0 < 0.65$.  Brighter stars were given higher priority.
 
Slitmasks from the 2010 observations were designed from SDSS $ugriz$
photometry \citep{SDSS_DR7}.  Stars were selected to lie within a
color range around a 14.1~Gyr, ${\rm [Fe/H]} = -1.63$, Padova
isochrone \citep{Girardi2004}.  The allowed range was 0.4~mag bluer
and 0.3~mag redder in $(g-r)_0$ color.  We also selected HB stars from
a box in the CMD: $21 > r_0 > 20$ and $-0.4 < (g-r)_0 < 0.0$.  As for
the 2009 observations, brighter stars were given higher priorities for
spectroscopic selection.

Because there is no published photometric data set for Ursa Minor that
covers the full extent of the galaxy and is sufficiently deep for our
purposes, the spectroscopic target selection for the 2012 observing
run relied on a number of different sources.

UMi has been observed by the Sloan Digital Sky Survey (SDSS), but the
SDSS DR7 \citep{SDSS_DR7} coverage nearly bisects the galaxy
along its major axis, with the southeast half of the galaxy included
but no data in the northwest half.  SDSS DR8 \citep{SDSS_DR8}, in
contrast, contains several stripes crossing the galaxy from southeast
to northwest, with $\sim20\arcmin$ gaps between each stripe.  The
difference in coverage between DR7 and DR8 is a result of different
data quality criteria in the SDSS processing of those data sets
(N. Padmanabhan 2012, private communication), but there is no evidence
for systematic photometric errors in either the DR7 or DR8 imaging in
this region.  We therefore generated a combined SDSS DR7 + DR8 catalog
for UMi, using DR7 measurements where available and DR8
elsewhere.\footnote{DR8 contained an astrometry error of up to
0.25\arcsec\ for northern targets \citep{SDSS_DR8_erratum, SDSS_DR9}, so for stars in the DR8 photometric catalog we used the
corrected positions provided in the early release of DR9 \citep{SDSS_DR9}.}

In addition to SDSS, we used the wide-field Washington/DDO51
photometry of \citet{Palma2003AJ....125.1352P}, the deeper VI imaging of \citet{Bellazzini2002AJ....124.3222B}
 in the center of the galaxy, and deep, wide-field gr
imaging covering 1~deg$^{2}$ with CFHT/Megacam from \citet{Munoz2018ApJ...860...65M}.  These catalogs were merged with the SDSS data taking
precedence, followed by stars in the Palma, Bellazzini, and Mu{\~
n}oz catalogs, in that order.  For the latter three data sets, we applied zero
point offsets to the astrometry so that the median position
differences with respect to SDSS DR9 of all stars in common were zero.

Spectroscopic targeting priorities for stars in the SDSS, Bellazzini,
and Mu{\~ n}oz data sets were determined using ${\rm [Fe/H]} = -2$,
10~Gyr isochrones from \citet{Dotter2008ApJS..178...89D}.  The RGB selection window
was defined so as to include all obvious members of UMi near the
center of the galaxy, with more generous color limits to the blue side
of the RGB to allow for unusually metal-poor stars.  We constructed a
similar selection window for horizontal branch (HB) stars by
generating a large number of synthetic HB stars with the online code
provided by \citet{Dotter2007} using the same age and metallicity as
for the RGB and then fitting a polynomial to determine the luminosity
of the HB as a function of color.  We assigned RGB candidates higher
priorities than HB candidates, with relative priorities determined by
magnitude within each category (where preference is given to brighter
stars), and then added priority bonuses for stars already confirmed to
be UMi members by \citet{Palma2003AJ....125.1352P} or the 2009/2010 Kirby data sets
described above.  Stars located within the bounds of either of the two
possible substructures in UMi identified by \citet{Pace2014MNRAS.442.1718P} were
given the highest priorities, and then slitmasks were placed to ensure
full coverage of both substructures.

\subsection{Reductions \& Measurements}

We reduced the DEIMOS data using the \texttt{spec2d} pipeline
developed by the DEEP2 team \citep{Cooper2012,Newman2013}.  The pipeline
cuts out the spectrally dispersed image of each slit from the raw
data.  The image is flat fielded, and a two-dimensional wavelength
solution is calculated from an exposure of Kr, Ne, Ar, and Xe arc
lamps.  
 The
typical root mean square difference between the known arc line wavelengths and the
calculated wavelength solution is 0.015~\AA\@ ($0.5 \kms$).
The stellar spectrum is extracted with ``optimal'' extraction \citep{Horne1986PASP...98..609H} and made into a
sky-subtracted, wavelength-calibrated, one-dimensional spectrum.  We
made some improvements to the pipeline appropriate for our purposes.
For example, the procedure for defining the extraction window was
optimized for extracting unresolved stars rather than extended
galaxies \citep{Simon2007ApJ...670..313S}.
The one-dimensional wavelength arrays were modified with slight offsets in order to align sky emission lines with their known wavelengths.  The wavelength arrays were also modified to remove the effect of differential atmospheric refraction perpendicular to the slit.

We measured radial velocities for each star by comparing the spectra
with a set of template spectra measured with DEIMOS\@ \citep[observed by][]{Kirby2015ApJ...810...56K}.  The velocities were calculated from the
wavelength shift in log space that minimized $\chi^2$ between the
target and template spectra.  This procedure is similar to a
cross-correlation \citep{Tonry1979}, but it also takes into account
variance in the observed spectrum.

We checked each radial velocity measurement by plotting the template
spectrum on top of the target spectrum shifted into the rest frame.
In several cases, the velocity measurement clearly failed, and the
spectra did not line up.  The typical cause was an artifact at the
edge of the target spectrum.  In these cases, we masked out the
offending region of the spectrum and repeated the velocity
measurement.

Because DEIMOS is a slit  spectrograph, mis-centering of stars can cause spurious offsets in the wavelength solution, which translate into offsets in the measured radial velocity.  We treated this offset as a shift in the zeropoint of the radial velocity.  We measured the zeropoint by using the observed wavelengths of telluric absoprtion from the Earth's atmosphere, which should be at rest in the geocentric frame.  This is sometimes known as the ``A-band correction'' \citep{Sohn2007ApJ...663..960S}.  We cross-correlated each of the observed spectra with the spectrum of a hot star observed with DEIMOS by \citet{Kirby2015ApJ...810...56K}.  The velocity zeropoint was taken to be the velocity shift required to align the template spectrum with the telluric absorption features.  This zeropoint was added to the radial velocity measured from the stellar absorption lines.  We inspected each spectrum to ensure that the telluric cross-correlation was valid, just as we did for the stellar absorption.  In about a dozen cases where the DEIMOS chip gap fell in the A-band, we had to re-evaluate the telluric cross-correlation after excluding some pixels around the chip gap.

We calculated velocity errors by resampling the target spectrum 1000
times.  In each Monte Carlo trial, we constructed a new spectrum by
perturbing the original flux value.  The magnitude of the perturbation
was sampled from a Gaussian random distribution with a variance equal
to the variance estimated for that pixel by \texttt{spec2d}.  The
velocity error was equal to the standard deviation of all of the Monte
Carlo trials.  \citet{Simon2007ApJ...670..313S} found from repeat measurements of the
same stars that this statistical error was an incomplete description
of the total error.  Following their example, we calculated the total
error by adding a systematic error of 2.2~km~s$^{-1}$ in quadrature
with the Monte Carlo statistical error.

We measured metallicities by comparing the continuum-normalized
observed spectra to a grid of synthetic spectra.  This procedure is
identical to that of \citet{Kirby2008, Kirby2010ApJS..191..352K}.  We started with a guess
at the effective temperature, surface gravity, and metallicity of the
star by combining the stars' colors and magnitudes with theoretical
isochrones.  The temperature and metallicity were allowed to vary to
minimize $\chi^2$ between the observed spectrum and the synthetic
grid.  Our measured value of [Fe/H]\footnote{${\rm [Fe/H]} = \log
  \frac{n({\rm Fe})/n({\rm H})}{n_{\sun}({\rm Fe})/n_{\sun}({\rm H})}$
  where $n$ is atomic number density.} is the one that minimized
$\chi^2$.

Errors on [Fe/H] were estimated from the diagonal terms of the
covariance matrix.  This is an incomplete estimate of the error,
largely due to covariance with temperature.  \citet{Kirby2010ApJS..191..352K} found
that adding a systematic error of 0.11~dex in quadrature with the
statistical error is an adequate estimate of the error.  We adopted
the same approach.

Our Keck/DEIMOS measurements are listed in Table~\ref{table:data}.

\input{table_data.tex}

\subsection{Validation}
\label{section:validation}

\begin{figure*}
\includegraphics[width=\textwidth]{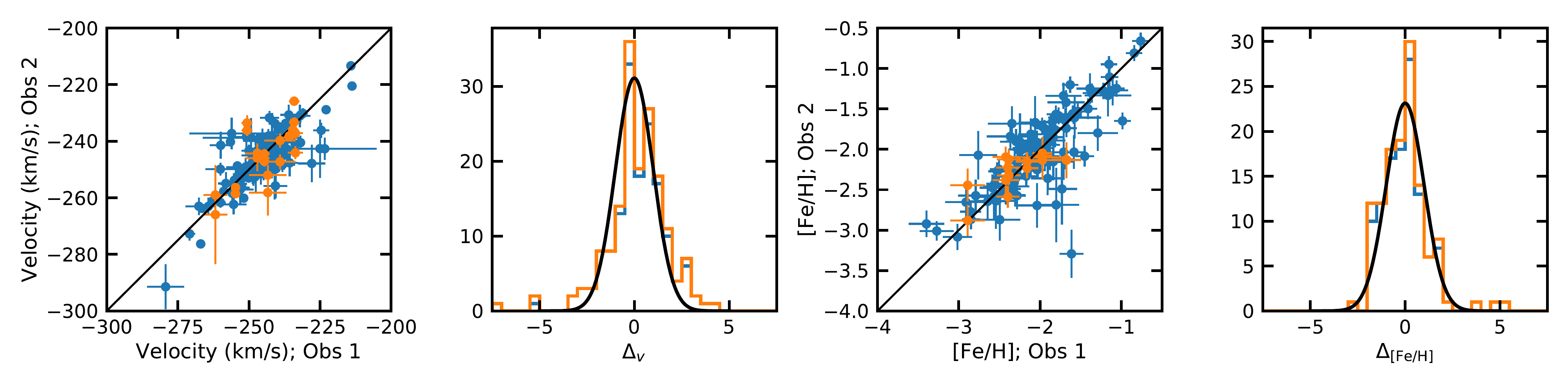}
\caption{Comparison of velocity (left panels) and [Fe/H] (right panels) measurements for stars with multiple measurements. Stars in blue (orange) have 2 (3) observations. The left and right velocity panels show the one-to-one comparison and the normalized difference between repeated observations (the left and right [Fe/H] panels are similar). 
Repeated measurements follow the overlaid normal distribution (mean of zero and variance of one).  }
\label{fig:repeat_stars_keck}
\end{figure*}

To verify our measurements we examined stars with repeated measurements (using a cross-match radius of 1\arcsec).
We find 155 stars with two spectral measurements and 12 stars with three measurements.
In Figure~\ref{fig:repeat_stars_keck}, we compare the repeated line-of-sight velocity and metallicity measurements.  
We compute the normalized difference, $\Delta$, between measurements.  For radial velocity measurements, $\Delta_{v}=(v_1-v_2) / \sqrt{\sigma_{\epsilon, v,1}^2+\sigma_{\epsilon, v,2}^2}$,  where $v$ and $\sigma_{\epsilon,v}$ correspond to the radial velocity and velocity error, respectively. 
If the  repeated measurements are consistent, the $\Delta$ distribution will follow a Gaussian distribution with a mean of zero and variance of unity.
Based on a Shapiro-Wilk test we find that our repeat measurements are consistent with a Gaussian distribution ($p=0.34$). 
After removing stars with clear velocity variation we find, $\overline{\Delta_{v}}=0.24$ and $\sigma_{\Delta_{v} }=1.18$ for repeated velocity measurements and $\overline{\Delta_{v}}=-0.15$ and $\sigma_{\Delta_{v} }=1.00$ for repeated metallicity measurements. 
The second and fourth panels of Figure~\ref{fig:repeat_stars_keck} show $\Delta$ for the velocity and metallicity measurements, respectively.
The tails in the velocity distribution could be due to unresolved binary stars.

To combine velocity measurements we use the weighted mean as the combined radial velocity.
For the  error we compute the variance of the weighted mean and weighted standard deviation and take the larger of the two for the combined velocity error.  
When the weighted standard deviation is larger than the variance of the weighted mean, the star may be variable in velocity and we use the weighted standard deviation to be conservative.
For metallicity measurements we use the weighted mean and variance of the weighted mean for all combined measurements.

\begin{figure*}
\includegraphics[width=\textwidth]{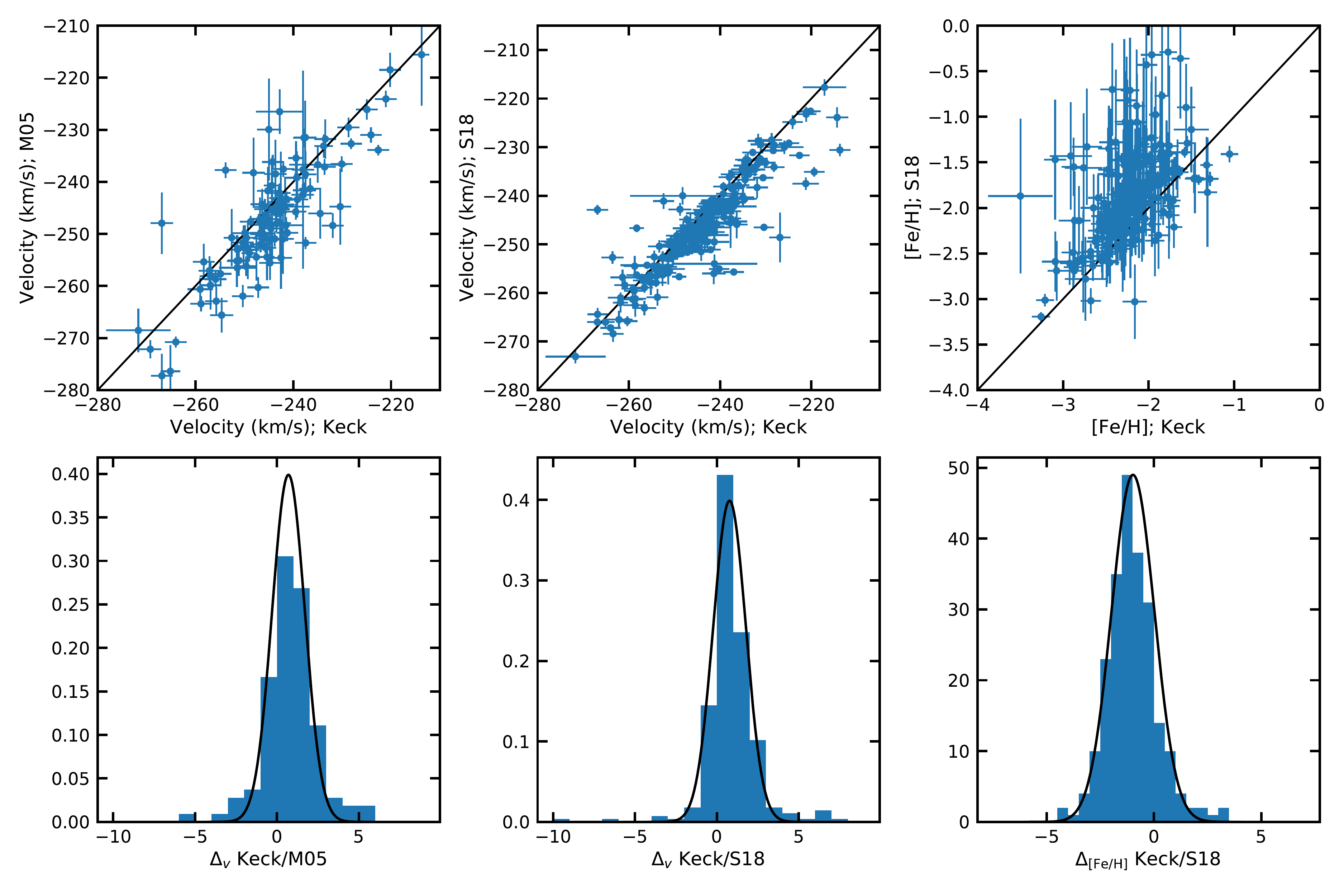}
\caption{Top: comparison of our velocity and metallicity measurements to previous works; M05 \citep{Munoz2005ApJ...631L.137M} and  S18 \citep{Spencer2018AJ....156..257S}.  
Bottom: normalized velocity ($\Delta_v$) and metallicity ($\Delta_{[\rm Fe/H]}$) differences between our analysis and previous studies. Overlaid are  normal distributions with different means ($\overline{\Delta}=+0.70, +0.77, -0.98$) but the same dispersion ($\sigma_{\Delta}=1$). The velocity and metallicity measurements have zero-point offsets between the different studies. 
All three distributions are consistent with a spread of one indicating that the relative errors are consistent between studies.}
\label{fig:repeats_compare}
\end{figure*}

We compare our velocity and metallicity measurements to two other large UMi spectroscopic samples (Figure~\ref{fig:repeats_compare}).  \citet{Munoz2005ApJ...631L.137M} presented velocity measurements from a combination of Keck/HIRES and WHT/WYFFOS \citep[the latter were originally presented in][]{Wilkinson2004ApJ...611L..21W}.
Previous analysis of this data set motivated our target selection \citep{Pace2014MNRAS.442.1718P}.
\citet{Spencer2018AJ....156..257S} utilized  MMT/Hectochelle to measure velocities and metallicities in UMi.  
There are 108 and 277 stars in common with our data set and \citet{Munoz2005ApJ...631L.137M} and \citet{Spencer2018AJ....156..257S}, respectively.
Overall, we find that our velocities are offset from both studies; the average mean normalized offset (after removing outliers) is $\overline{\Delta_v}=0.70$, $\sigma_{\Delta_v}=1.17$ and  $\overline{\Delta_v}=0.77$,  $\sigma_{\Delta_v}=0.92$ for \citet{Munoz2005ApJ...631L.137M} and \citet{Spencer2018AJ....156..257S},  respectively.  
These offsets are likely caused by zero-point offsets in the radial velocity templates  assumed between different analyses. 
We find that there is an offset in the metallicity between the \citet{Spencer2018AJ....156..257S}  MMT/Hectochelle measurements and find $\overline{\Delta_{\rm [Fe/H]}}=-0.98$ and $\sigma_{\Delta_{\rm [Fe/H]}}=1.03$.  
This offset may be due to the differences in measurement techniques or due to the different spectral ranges and resolutions.  
Based on our UMi analysis, we find these normalized offsets translate to offsets of $\Delta\overline{v}\approx2.4\kms$ and $\Delta \overline{\rm [Fe/H]}\approx0.16$.
While there are  offsets between the different studies, we find that the errors are consistent between the different studies.

\subsection{Final Catalog Selection}
\label{section:catalog}

\begin{figure*}
\includegraphics[width=\textwidth]{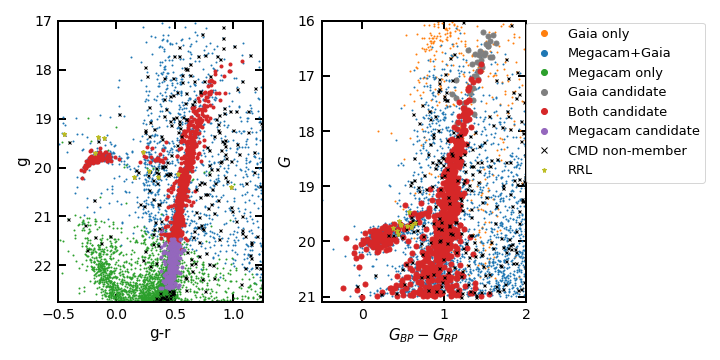}
\caption{Color-magnitude diagrams in g-r vs g (Megacam; left) and $G_{BP}-G_{RP}$ vs $G$ ({\it Gaia}; right) of UMi. Blue, orange, and green    points show stars that are contained in both catalogs, exclusive to the {\it Gaia} catalog, or exclusive to the Megacam catalog.  Gray, red, and purple stars pass our color-magnitude selection and are candidate UMi members; gray points utilize {\it Gaia} bands while the reminder use Megacam photometry.  Olive stars are RR Lyrae stars identified in PS1 or {\it Gaia} catalogs.  Stars excluded due to their location on the color-magnitude diagram  are shown in black.
}
\label{fig:cmd}
\end{figure*}

As mentioned in Section~\ref{section:obs}, at the time of our target selection and observations there was not a deep and wide-field public photometric catalog for UMi.  
There are now several, including the wide-field CFHT/Megacam data \citep{Munoz2018ApJ...860...65M}, Pan-STARRS1 \citep{Chambers2016arXiv161205560C}, and {\it Gaia} DR2 \citep{GaiaBrown2018A&A...616A...1G}.
We use the gr CFHT/Megacam photometry \citep{Munoz2018ApJ...860...65M} for the majority of the sample  and {\it Gaia} DR2 photometry ($G$ and  $G_{\rm BP}-G_{\rm RP}$) for the brightest stars in the sample that are  saturated in the Megacam catalog ($G < 18$, roughly $g\sim18.75$).  
We perform a broad isochrone selection  to pick stars with colors and magnitudes consistent with UMi.  
For the {\it Gaia} DR2 photometry, we base on cuts on the selection by \citet{GaiaHelmi2018A&A...616A..12G}.
For the Megacam photometry and stars on the red giant branch, we select stars within a color window of 0.12 mag from an ${\rm [Fe/H]} = -2$, 10~Gyr isochrone  \citep{Dotter2008ApJS..178...89D} and exclude stars fainter than $g=22.5$.
For the horizontal branch selection, we select stars within magnitude windows of 0.3 mag and 0.2 mag for stars with $g-r<0$ and $0<g-r<0.55$ respectively.  The  window in the red horizontal branch region is narrower  due to the increase in MW interlopers. 
This selection is similar  to the target selection on the preliminary Megacam photometry.

We  use {\it Gaia} DR2 proper motions to improve MW foreground identification and determine the proper motion of UMi.
We cross match our spectroscopic sample with the {\it Gaia} DR2 catalog with a cross-match radius of 1\arcsec\ (for most stars the cross-match radius is less than 0\farcs5) and find  1060 matches out of 1532 stars with  an astrometric solution (\mbox{\tt astrometric\_params\_solved}=31). 
We utilize the stellar parallax to identify nearby disk stars. We consider all stars with a non-zero parallax as members of the MW disk ($\varpi - 3\sigma_{\varpi}>0$). 
We calculate the tangential velocity for each star assuming it is at the distance to UMi (Table~\ref{table:UMi_properties}).  Any star that would be unbound at UMi's distance ($v_{\rm tan} - 3\sigma_{v_{\rm tan}} > v_{\rm escape}$) is considered a nearby MW foreground star \citep{Pace2019ApJ...875...77P}.

RR Lyrae stars are variable in velocity and not suited for kinematic analysis. 
We cross-match our sample to the Pan-STARRS RR Lyrae  \citep{Sesar2017AJ....153..204S} and {\it Gaia} DR2 RR Lyrae catalogs \citep{Clementini2019A&A...622A..60C} and  find 17 and 15 matches, respectively  (with a 0\farcs5 cross-match radius). 
We find a total of 18 RR Lyrae in our spectroscopic sample as 12 stars overlap between the RR Lyrae catalogs.
The velocities and locations on a color-magnitude diagram of all 18 stars are consistent with membership in UMi.  
Two RR Lyrae each have two velocity measurements and exhibit clear  velocity variation ({\it Gaia} source\_id, 1645449593695899264: $-298.3\pm2.5$ and $-279.3\pm2.4$ and 1645468079235094784: $-229.6\pm2.6$ and  $-235.7\pm2.5$).
We exclude all known RR Lyrae from our analysis.

We use the surface gravity sensitive spectral feature at  Na {\rm \scriptsize I} at 8190 \AA $\;$  to identify additional MW foreground stars \citep{Spinrad1971ApJS...22..445S, Cohen1978ApJ...221..788C}.
Stars with $\sigma_{\epsilon, v}>20 \kms$ are excluded and we do not use the metallicity measurements for stars with  $\sigma_{\epsilon, {\rm [Fe/H]}}>0.5$~dex.
We compute the $\chi^2$ of a  non-variable velocity for each star  (i.e., any variation is due just to measurement errors) and  the corresponding p-value.
We exclude seven stars with clear indications of velocity variability from our analysis ($\Delta_v>3$ and/or $p <0.01$).
After our  color-magnitude selection and removing stars with indications of variability we have 1009 candidate UMi stars. Based on the parallax, large proper motion, or Na {\rm \scriptsize I} doublet, 64 of these stars are immediately identified as MW foreground stars.

\section{Methodology}
\label{section:method}

To identify dSph members and disentangle chemodynamical populations, we construct mixture models and assess statistical significance with model selection tests. 
This analysis builds upon the statistical framework of \citet{Walker2011ApJ...742...20W} by extending both the stellar distribution and selection function to axisymmetric systems, by including a  Milky Way model, and by including proper motion to significantly improve MW foreground selection.
To address the significance of additional populations we compute the Bayes' Factor between single and multi-population models.

We work in a Bayesian framework for disentangling different  components.
The probability of observing a data set, $\bm{x}=\lbrace\bm{x}_i \rbrace$, assuming a particular hypothesis or model, H, characterized by parameters, $\mathscr{M}$, is given by the likelihood: $\mathcal{L}(\bm{x}|\mathscr{M}) = \mathcal{P}(\bm{x} | \mathscr{M}, H) $.
We are interested in solving for the model parameters, found by determining the posterior distribution,  $\mathcal{P}(\mathscr{M}| \bm{x}, H)$. 
The posterior and likelihood are related via Bayes' Theorem:

\begin{equation}
\mathcal{P}(\mathscr{M}| \bm{x}, H) = \frac{\mathcal{L}(\bm{x}|\mathscr{M}) Pr(\mathscr{M},H) }{ \mathcal{P}(\bm{x}, H)},
\end{equation}

\noindent where $Pr(\mathscr{M},H)$, is the prior distribution representing any previously known information about the model under consideration and $\mathcal{P}(\bm{x}, H)$, is the marginal likelihood, a normalizing factor.  
The marginal likelihood is commonly referred to as the Bayesian Evidence in Astrophysics.  It is given by:

\begin{equation}
Z = \mathcal{P}(\bm{x}, H) = \int_{\mathscr{M}} \mathcal{P}(\bm{x} | \mathscr{M},H) Pr(\mathscr{M}) d\mathscr{M}.
\label{eq:bayes_evidence}
\end{equation}

\noindent  For general parameter estimation, computing the normalization is unnecessary. However, it is useful for model selection purposes.
To evaluate the posterior distribution and evidence we utilize   Multimodal Nested Sampling  \citep{Skilling2004, Feroz2008MNRAS.384..449F, Feroz2009MNRAS.398.1601F}. 
The nested sampling algorithm transforms the multi-dimensional evidence integral (Equation~\ref{eq:bayes_evidence}) into a one-dimensional integral over the `prior volume.'  The integral is evaluated by sampling the likelihood in a decreasing sequence of prior volumes, assuming that the inverse of the prior volume exists and is a monotonically decreasing function.
As a by-product of sampling the likelihood, the posterior is also computed  \citep[for a more detailed description, see][]{Feroz2008MNRAS.384..449F, Feroz2009MNRAS.398.1601F}.

For our analysis, the likelihood at each data point is independent and therefore, the total likelihood is the product of the likelihood at each data point,

\begin{equation}
\mathcal{L}(\bm{x}|\mathscr{M}) = \prod_{x=i}^{N} \mathcal{L}(\bm{x}_i|\mathscr{M}).
\end{equation}

\noindent The likelihood at each data point is a mixture of a Milky Way (MW) foreground and a dSph population \citep[e.g.][]{Koposov2011ApJ...736..146K,  Martinez2011ApJ...738...55M, Walker2011ApJ...742...20W}:   

\begin{equation}
\mathcal{L}(\bm{x}_i|\mathscr{M})  = f_{{\rm MW}} \mathcal{L}_{{\rm MW}}(\bm{x}_i|\mathscr{M}_{{\rm MW}}) + f_{\rm dSph} \mathcal{L}_{\rm dSph} (\bm{x}_i|\mathscr{M}_{\rm dSph}),
\label{eq:mixture}
\end{equation}

\noindent where $f_{\rm MW/dSph}$ denotes the observed fraction of stars within that component  and $f_{\rm MW}+f_{\rm dSph} = 1$.  
In general, additional components can be added with the constraint: $\sum_c f_c=1$.
This may include additional dSph components  \citep[e.g.][]{Amorisco2012ApJ...756L...2A, Kordopatis2016MNRAS.457.1299K} or additional foreground components \citep[for example, the background model for M31 satellites is composed of stars from both the MW and M31 halo,][]{Tollerud2012ApJ...752...45T,  Collins2013ApJ...768..172C, Gilbert2018ApJ...852..128G}.  
For brevity, we will drop the parameter denotation ($|\mathscr{M}$) from the likelihood arguments.
In some of the later analysis, to identify chemodynamic components we decompose the dSph likelihood into  multiple components, 

\begin{equation}
\mathcal{L}_{\rm dSph}(\bm{x}_i)  = f_1 \mathcal{L}_1 (\bm{x}_i) + f_2  \mathcal{L}_2 (\bm{x}_i)\,.
\label{eq:dsph_mix}
\end{equation}

To  determine a star's membership in a component  we compute the ratio of the component likelihood to the total likelihood for that star \citep[e.g.,][]{Martinez2011ApJ...738...55M, Pace2014MNRAS.442.1718P}.  
In more concrete terms, the membership probability for the $i$th star to be in component $c$ is:

\begin{equation}
p_{c, i} = \frac{f_c \mathcal{L}_c (\bm{x}_i)}{\sum_k f_{k} \mathcal{L}_{k} (\bm{x}_i)}.
\label{eq:member}
\end{equation}

\noindent We compute the membership from the posterior distribution.
Each star will have a probability distribution of membership for each component, for practicality, we use the median membership for derived quantities.

\noindent{\bf Model Selection}--To determine whether multiple components are significant we compute the logarithmic Bayes' factor, $\ln{B}$.  
The Bayes' factor compares the relative odds in favoring model A over model B after examining the data.
It is the ratio of the evidences (Equation~\ref{eq:bayes_evidence}) computed for each model (with the assumption that apriori both models are equally favored),

\begin{equation}
B_{AB}  = \frac{\mathcal{P}(x, H_A)}{\mathcal{P}(x, H_B)} = \frac{Z_A}{Z_B}.
\end{equation} 

\noindent It naturally incorporates Occam's razor as larger or more complicated model spaces are penalized. 
For models $A$ and $B$, $\ln{B_{AB}} = \ln{B}>0$ favors model $A$ and $\ln{B}<0$ favors model $B$.  
To interpret the significance of the Bayes' factor, we follow the empirical `Jeffreys' scale.' 
The  ranges  of $<1, 1-2.5, 2.5-5, >5$ correspond to regions of  inconclusive, weak, moderate, and strong evidence respectively \citep[see][]{Trotta2008ConPh..49...71T}. 

\subsection{Selection Function}
\label{section:sf}

\begin{figure}
\includegraphics[width=\columnwidth]{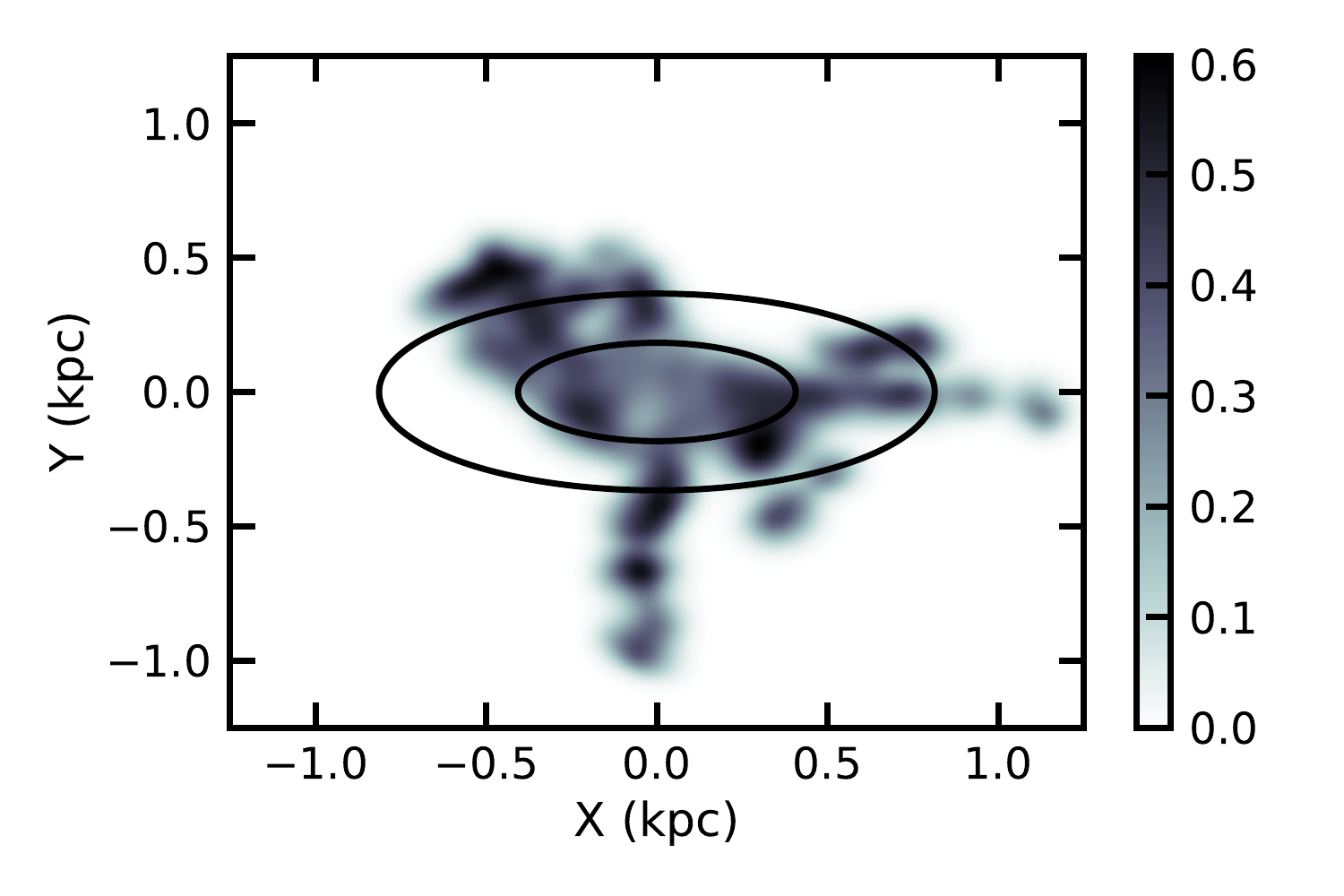}
\caption{Selection function of Keck/DEIMOS sample  (smoothing scale of k=50 pc).  For reference, we have overlaid ellipses representing the one and two times the half-light radius.  The x and y axes are aligned with the major and minor axes of UMi, respectively (position angle of 50\degree).
}
\label{fig:sf}
\end{figure}

Due to limited telescope time,  not all spectroscopic candidates can be observed.  
The spatial distribution of stars with spectroscopic measurements generally does not follow the intrinsic spatial distribution of stars (due to telescope field-of-view, mask size, etc). 
To ensure that spatial parameters of the dSph can be recovered from the spectroscopic distribution we compute the selection function,  $S(x,y)$.  The selection function acts as a  mapping between the observed and intrinsic spatial distribution of stars \citep{Wang2005, Walker2011ApJ...742...20W}. 

To construct $S(x,y)$, we  smooth the ratio of observed to candidate stars within the UMi region \citep{Walker2011ApJ...742...20W}:

\begin{equation}
S(x,y) = \frac{d\mathrm{ N_{obs}} (x,y)}{d \mathrm{N_{cand}}(x,y)} \approx \frac{\sum_{i=1}^{\mathrm{N_{obs}}} exp{ \left[ - \frac{1}{2} \frac{(x_i - x)^2 + (y_i - y)^2}{k^2} \right] }}{\sum_{i=1}^{\mathrm{N_{cand}}}  exp{ \left[ - \frac{1}{2} \frac{(x_i - x)^2 + (y_i - y)^2}{k^2} \right] }} \,,
\label{eq:sf}
\end{equation}

\noindent where $k$ is the smoothing scale \citep{Walker2011ApJ...742...20W}.
$N_{\mathrm{cand}}$ are all stars in the UMi region that fall within our photometric selection in Section~\ref{section:catalog} while $N_{{\rm obs}}$ are all spectroscopically observed stars within the same photometric selection.  
We use the projected spatial positions ($x$,$y$) instead of the projected radial positions ($R=\sqrt{x^2+y^2}$) because UMi is more aspherical than Fornax and Sculptor \citep[$q\approx0.45$ compared to $q\approx0.71, 0.67$;][]{Munoz2018ApJ...860...66M}.  
In addition, the stellar populations may not necessarily have the same ellipticity.

To set $k$, we construct and observe mock data sets with different values of $k$ ranging from 25 to 400 parsecs (1$\arcmin$ to 18$\arcmin$ at $d=76$ kpc).  We find, that for spherically symmetric systems, the choice of $k$, does not make an appreciable difference\footnote{\citet{Walker2011ApJ...742...20W} reach the same conclusion.}.
Whereas, for axisymmetric systems, an incorrect choice in $k$ will strongly bias the recovered structural parameters.
Our tests, with a layout approximating the locations and sizes of Keck/DEIMOS masks, show that $50 \,\mathrm{pc} \leq k \leq 75 \,\mathrm{pc}$  ($2\arcmin \leq k \leq 3\arcmin $) correctly recovers the input  structural parameters.  A larger choice of $k$ will bias the structural posteriors.  The spatial scale will be underestimated and axis ratio overestimated.  The spatial bias  increases as $k$ increases.

In Figure~\ref{fig:sf}, we plot the selection function we use for the UMi analysis with $k=50$ pc.  We align the x-axis and the major axis    \citep[using a position angle of $\theta=50\degree$][]{Munoz2018ApJ...860...66M}.   We fix $k=50$ pc for the main analysis. 

\subsection{Likelihoods}
\label{section:likelihood}

This analysis uses line-of-sight velocity ($v$, $\sigma_{\epsilon,v}$), metallicity (${\rm [Fe/H]}$, $\sigma_{\epsilon,\rm [Fe/H]}$), spatial position ($x$, $y$),  and proper motion\footnote{$\mu_{\alpha*}=\mu_{\alpha}\cos{\delta} = -\mu_{W}$ } ($\mu_{\alpha*}$,  $\sigma_{\epsilon, \mu_{\alpha}*}$, $\mu_{\delta}$, $\sigma_{\epsilon,\mu_{\delta}}$, $C_{\mu_{\alpha*} \times \mu_{\delta}}$) for each data point\footnote{We note that not all stars have metallicity or proper motion measurements.  For each data point without a particular measurement, that likelihood term must be integrated out (i.e. all possible metallicity or proper motion values are considered). As the individual  likelihood terms are normalized to unity, integrating over all possible  values drops the term from the likelihood and the membership probability for that star will only consider spatial and velocity information. }.  
Our analysis utilizes synthetic measurements of the iron abundance, [Fe/H].  Other works have utilized different metallicity tracers, for example, the Ca {\sc ii} triplet \citep{Battaglia2008ApJ...681L..13B}, or the $\Sigma$Mg index \citep{Walker2011ApJ...742...20W}. 

We assume that probability distributions of the velocity, position, metallicity, and proper motions  are independent of one another, therefore, the likelihood of the dSph or MW is, 

\begin{equation}
\mathcal{L}_{\rm dSph/MW}= \mathcal{P}^{{\rm vel}} \times \mathcal{P}^{{\rm spatial}} \times\mathcal{P}^{{\rm [Fe/H]}} \times\mathcal{P}^{{\rm PM}} \,.
\end{equation}

The majority of the probability distributions  are assumed to be  Gaussian or multivariate Gaussian distributions. 
For example, the probability distribution of the velocity term is: $\mathcal{P}^{{\rm vel}} ({v_i, \sigma_{\epsilon,v}}_i | \overline{v}, \sigma_v) = \mathcal{N}(v_i-\overline{v}, \sqrt{\sigma_{\epsilon,v, i}^2 + \sigma_v^2})$, where $\mathcal{N}(a,b)$ is a normal distribution where the mean, a, and dispersion, b.  $\overline{v}$ and $\sigma_v$ are the  average velocity and velocity dispersion, respectively.

For an extended object, like a dSph, the average velocity is a function of spatial position due to projection effects, sometimes referred to as the perspective motion, and we replace $\overline{v}$ with $v_{\rm rel}(\alpha, \delta)$ \citep{Feast1961, van_der_Marel2002, Kaplinghat2008ApJ...682L..93K, Walker2008ApJ...688L..75W}.  
The effect can be understood as the difference between the $z$ coordinate and line-of-sight direction, for example, $v_{\rm rel}(x,y) \approx -v_z + v_x x/d + v_y y/d$ \citep{Kaplinghat2008ApJ...682L..93K}. 
Our implementation  of the  perspective motion follows the appendix of \citet{Walker2008ApJ...688L..75W}.
For UMi, the effect is $\Delta \lvert v\rvert \sim0.1-0.2 \kms$.

The bright dSphs have flat isothermal line-of-sight velocity dispersions \citep{Munoz2005ApJ...631L.137M, Munoz2006ApJ...650L..51M, Walker2007ApJ...667L..53W}, therefore we initially assume $\sigma_v$ is constant with radius.  
However, this assumption may not hold for individual chemodynamic components \citep{Battaglia2008ApJ...681L..13B, Strigari2017ApJ...838..123S}; we address this in Section~\ref{section:results}.  
We assume the metallicity likelihood is a Gaussian distribution with a free mean, $\overline{\rm [Fe/H]}$ and dispersion, $\sigma_{\rm [Fe/H]}$.

The likelihood for the spatial distribution is \citep{Wang2005, Walker2011ApJ...742...20W}, 

\begin{equation}
\mathcal{P}^{{\rm spatial}}(x_i, y_i) = \frac{S(x_i,y_i) \Sigma(x_i, y_i) }{\int S(x,y) \Sigma(x,y) dA}\,,
\end{equation}

\noindent where $\Sigma(x,y)$ is the projected stellar distribution and $S(x,y)$ is the selection function (Section~\ref{section:sf}) 
The denominator ensures that the positional likelihood is normalized and acts as a weight for spatial profile reconstruction. 
We model the projected stellar distribution with an elliptical Plummer profile \citep{Plummer1911MNRAS..71..460P},

\begin{equation}
\Sigma(x, y) =  \frac{1}{(1 - \epsilon) \pi r_p^2 } \frac{1}{ \left( 1 + R_e^2/r_p^2  \right)^{2} } \,,
\label{eq:plummer}
\end{equation}

\noindent where $R_e= (x^2 + y^2 /(1 - \epsilon)^2)$ is the elliptical radius, $r_p$ is the stellar scale radius, and $q$ is the axis ratio and $\epsilon$ is the ellipticity.   
For this analysis we will approximate the spherically averaged half-light radius as $R_h=r_p \sqrt{1-\epsilon}$.
We use a Plummer profile for simplicity but note that stellar distribution profiles with additional parameters may provide better fits \citep[e.g., S\'{e}rsic;  ][]{Munoz2018ApJ...860...66M}

We model the proper motion likelihood as a multivariate Gaussian  due to the correlated error between the $\mu_{\alpha*}$ and $\mu_\delta$ components \citep{Pace2019ApJ...875...77P}.
The proper motion likelihoods have free means, $\overline{\mu_{\alpha*}}$ and $\overline{\mu_{\delta}}$, and dispersions, $\sigma_{\mu_{\alpha*}}$ and $\sigma_{\mu_{\delta}}$.  For the dSph component we fix $\sigma_{\mu_{\alpha*}} = \sigma_{\mu_{\delta}} = 10 \kms = 0.03 \, {\rm mas \, yr^{-1}}$ and leave the MW dispersions as free parameters.  The proper motion errors for the brightest UMi stars are $\approx40\kms$ and rapidly increase  for fainter stars. 
Due to the large errors it is not possible to infer the  intrinsic dSph proper motion dispersions but this may be possible in future {\it Gaia} data releases.
The expected MW halo dispersion will be $\approx100-200 \kms$ depending on the component (e.g., disk, halo).

The individual MW probability distributions are similar to the dSph components.  We assume Gaussian or multivariate Gaussians for the velocity, metallicity, and proper motion components with free means ($\overline{v}_{\rm MW}$, $\overline{\rm [Fe/H]}_{\rm MW}$, $\overline{\mu_{\alpha *,\, {\rm MW}} }$, $\overline{\mu_{\delta,\, {\rm MW}} }$) and variances ($\sigma_{v, \, {\rm MW}}$, $\sigma_{\rm [Fe/H], \, MW}$, $\sigma_{\mu_{\alpha *},\, {\rm MW}}$, $\sigma_{\mu_{\delta },\, {\rm MW}}$).  For the spatial distribution we assume that it is uniform within the UMi region. 
After accounting for the selection function the spatial probability distribution is: $\mathcal{P}^{{\rm spatial}}_{\rm MW}(x_i, y_i) = S(x_i,y_i) / \int S(x,y)  dA $.

There are a number of stars that can be immediately identified as foreground stars due to the surface gravity sensitive spectral feature at  Na {\rm \scriptsize I} at 8190 \AA $\;$, a non-zero parallax, and/or an extremely large tangential velocity (see Section~\ref{section:catalog}).  We do not want to immediately remove these stars from the sample as their $v$ and $\feh$ will help construct the MW foreground distribution and assist in identifying MW stars without {\it Gaia} astrometric measurements.  
These stars are only used in the MW likelihood and we do not include their proper motion information in the likelihood.

\section{Results}
\label{section:results}
\subsection{Ursa Minor Properties}

\begin{table}
 \caption{Properties of Ursa Minor from the literature and selected posterior values from our dSph and MW mixture model. $r_p$, $\epsilon$, $\theta$ assume a Plummer profile. $R_h = r_p \sqrt{1-\epsilon}$. }
 \label{table:UMi_properties}
 \begin{tabular}{lll}
  \hline
  Property & value & citation\\
  \hline
$\alpha_o$ (deg) & 227.2420  & \citep{Munoz2018ApJ...860...66M} \\ 
$\delta_o$ (deg) & 67.2221  & \citep{Munoz2018ApJ...860...66M} \\ 
$\theta$ (deg) &  $50\pm1$ & \citep{Munoz2018ApJ...860...66M} \\
$\epsilon$ & $0.55\pm0.01$ & \citep{Munoz2018ApJ...860...66M} \\
$r_p$ (arcmin) & $18.3\pm0.11$ &  \citep{Munoz2018ApJ...860...66M} \\
$R_h$ (pc) & $271\pm3$ &  \citep{Munoz2018ApJ...860...66M} \\
$M_V$  & $-9.03\pm0.05$ &  \citep{Munoz2018ApJ...860...66M} \\
d (kpc) & $76\pm5$ kpc & \citep{Bellazzini2002AJ....124.3222B} \\
\hline
$\overline{v}\,  (\kms)$ & $-244.7_{-0.3}^{+0.3}$ & This Work \\ 
$\sigma_{v}\,  (\kms)$ & $8.7_{-0.3}^{+0.3}$ & This Work \\ 
$\mu_{\alpha *} \, ({\rm mas \, yr^{-1}})$ & $-0.151_{-0.014}^{+0.014}$ & This Work \\
$\mu_{\delta} \, ({\rm mas \, yr^{-1}})$ & $0.065_{-0.013}^{+0.013}$ & This Work \\ 
$\overline{{\rm [Fe/H]} }$ & $-2.13_{-0.02}^{+0.02}$ & This Work \\ 
$\sigma_{\rm [Fe/H]}$ & $0.35_{-0.01}^{+0.01}$ & This Work \\ 
$\epsilon$ & $0.59_{-0.02}^{+0.02}$ & This Work \\ 
$r_p$ (pc)  & $447_{-20}^{+23}$ & This Work \\ 
 $R_h$ (pc)  & $287_{-10}^{+11}$ & This Work \\
  \hline  
 \end{tabular}
\end{table}

\input{table_properties.tex}

We first explore a model with a single dSph component to identify UMi and foreground MW stars.  
The center, position angle, and distance are fixed to literature measurements and the adopted values are listed in Table~\ref{table:UMi_properties}.
There are 1009 stars that pass our cuts and are used as input in the mixture model.  
Stars identified as MW stars based on their parallax, Na {\rm \scriptsize I} doublet, or large proper motions are fixed to the MW component and only their  velocity and metallicity information is included in the mixture model.

We find  $\overline{v}=-244.7 \pm 0.3 \kms$ and $\sigma_v=8.7\pm0.3\kms$ for  UMi.
The velocity dispersion is  consistent with  $\sigma_v=9.5\pm1.2 \kms$ \citep{Walker2009ApJ...704.1274W} and in between the measurements of  $\sigma_v=8.0\pm0.3\kms$ \citep{Spencer2018AJ....156..257S}\footnote{We note that in our analysis of the \citeauthor{Spencer2018AJ....156..257S} data, we find a larger $\sigma_v$ value that is consistent with our  Keck/DEIMOS results.  For further details see Section~\ref{section:mmt}.} and  $\sigma_v=11.5\pm0.6$  \citep{Munoz2005ApJ...631L.137M}. 
The \citet{Munoz2005ApJ...631L.137M} data set is more spatially extended and is a combination of WHT/WYFFOS \citep{Wilkinson2004ApJ...611L..21W} and Keck/HIRES observations.  
The difference in $\sigma_v$ could be due to a $\sigma_v$ that increases with radius, or  the \citet{Munoz2005ApJ...631L.137M} data set could consist of a larger fraction of stars in the kinematically hotter population (see next section). Another possibility is a velocity offset between the two instruments which inflates $\sigma_v$. 
The metallicity properties we find are $\fehavr=-2.13\pm0.02$ and $\sigma_{\feh}=0.35\pm0.01$ which agrees with $\fehavr=-2.13\pm0.01$ and $\sigma_{\feh}=0.34$ from \citet{Kirby2011ApJ...727...78K}.  
The spatial properties we derive assuming a Plummer distribution are  $r_{{\rm p}}=447_{-20}^{+23}\,{\rm pc}$,  $\epsilon=0.59\pm0.02$, and $R_{\rm h}=287_{-10}^{+11} \, {\rm pc}$ ($R_h = r_p \sqrt{1 - \epsilon}$).
This agrees with $q=0.45\pm0.01$ and $R_{h} = 271 \pm 3\,{\rm pc}$ derived from deep, wide-field photometric data   \citep{Munoz2018ApJ...860...66M}.
Our proper motion measurement agrees within the errors with previous {\it Gaia} DR2 measurements \citep{GaiaHelmi2018A&A...616A..12G, Fritz2018A&A...619A.103F}.

Of the 1009 candidate stars, 892 (898) have membership greater than $>0.95$ ($>0.90$).  
The inclusion of {\it Gaia} proper motions results in very few stars with intermediate membership; only 12 stars have membership values in the range $0.05 < p_{\rm dSph} < 0.90$ (only one of which has a {\it Gaia} DR2 astrometric solution).  
Overall, we find our  metallicity,  spatial, and proper motion properties of  UMi  are consistent  with previous work whereas our velocity dispersion  measurement is in moderate tension ( $\sim2-3\sigma$) with both \citet{Munoz2005ApJ...631L.137M} and \citet{Spencer2018AJ....156..257S} but lies in between these  measurements.

\subsection{Detection of Two Chemodynamical Stellar Populations}

\begin{figure}
\includegraphics[width=\columnwidth]{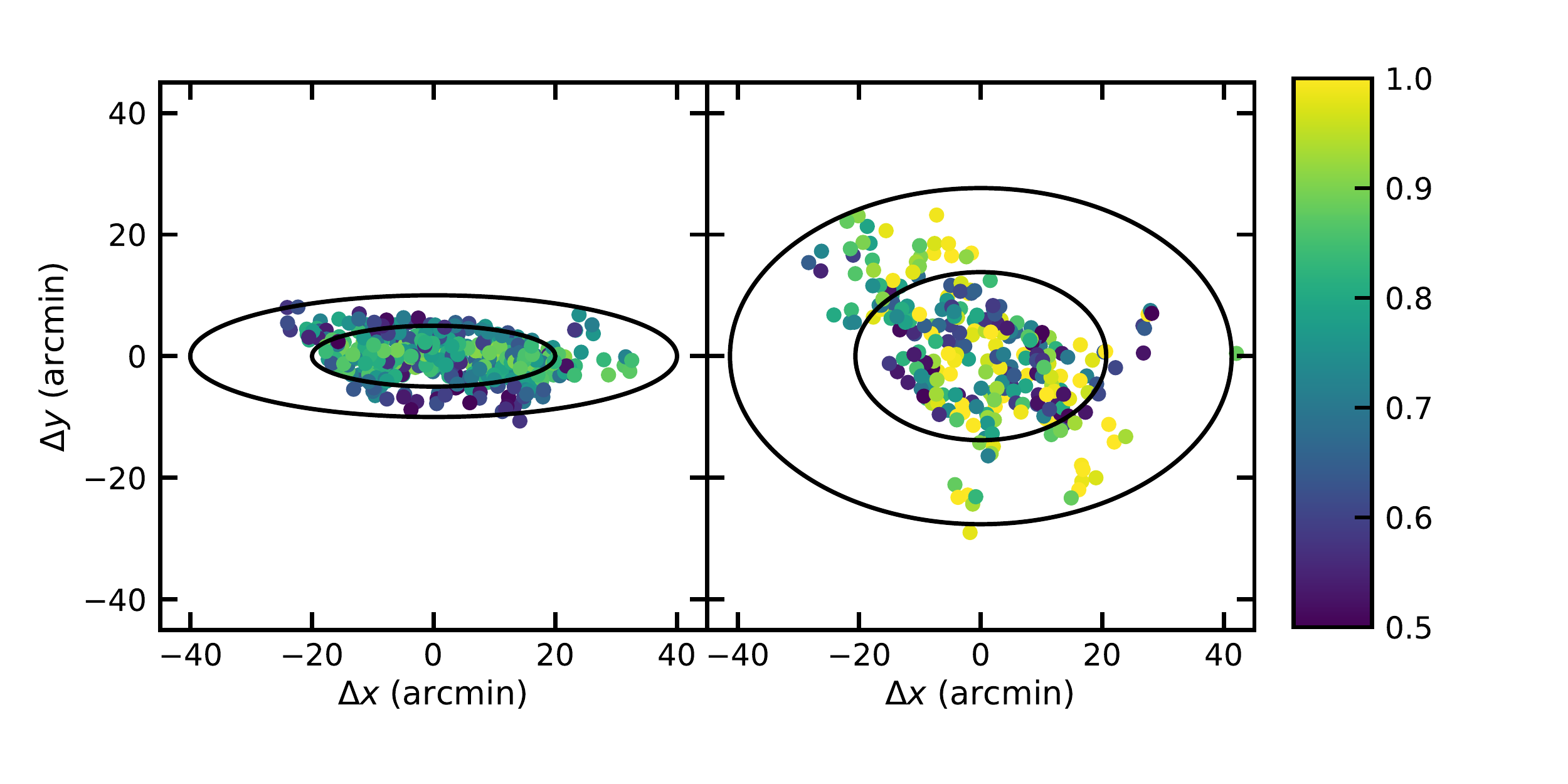}
\caption{Spatial distribution of stars in the metal-rich (left) and metal-poor (right) chemodynamic populations.  Stars are assigned based on the population in which they have larger membership probability. 
Ellipses with one and two times the half-light radius are shown.
The metal-rich population is significantly more elliptical than the metal-poor population.
}
\label{fig:ellipticity}
\end{figure}

\begin{figure}
\includegraphics[width=\columnwidth]{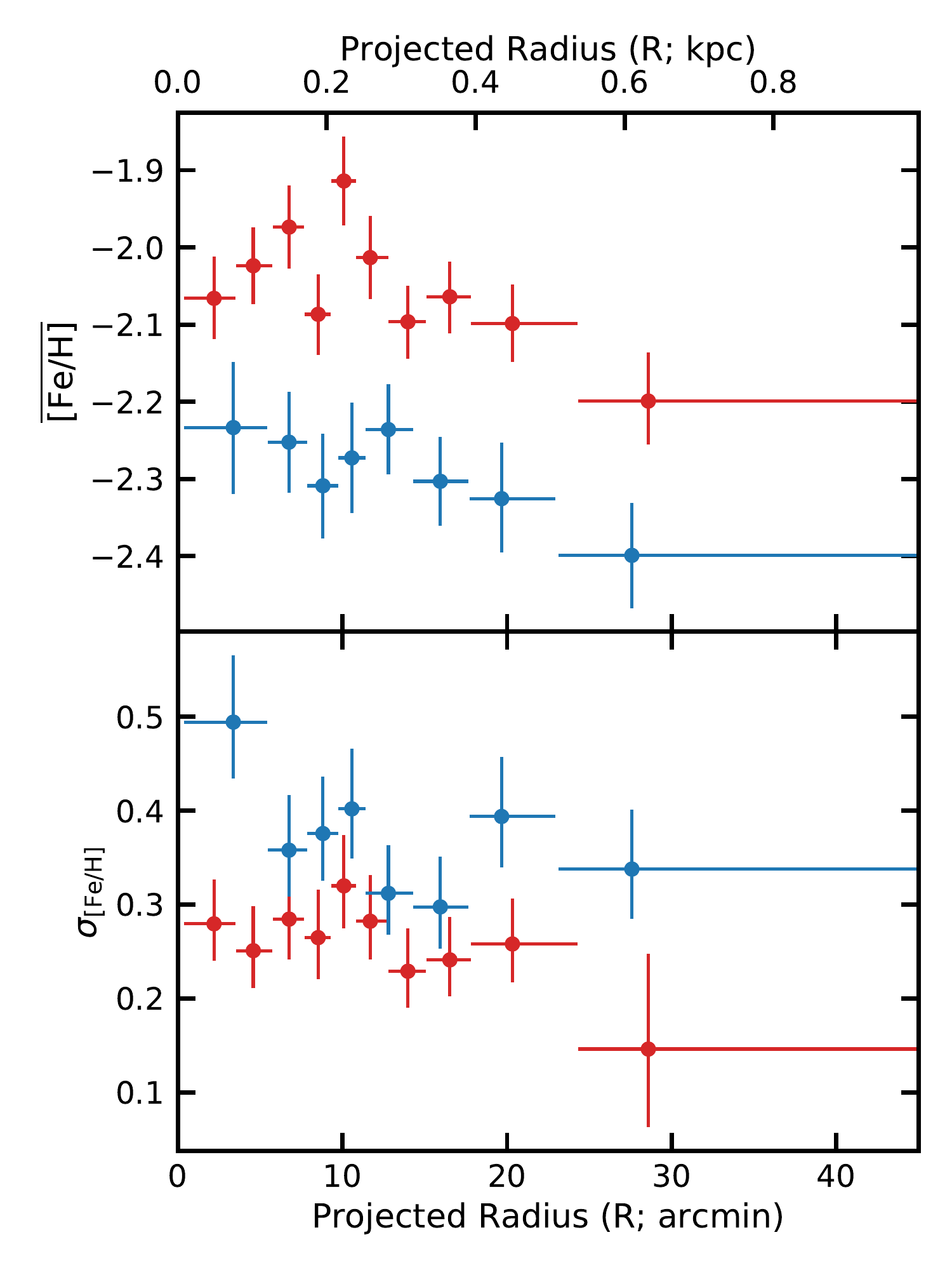}
\caption{Radial distribution of the mean metallicity (top; $\overline{\rm [Fe/H]}$) and metallicity dispersion (bottom; $\sigma_{\rm [Fe/H]}$) for the metal-rich (red) and metal-poor (blue) populations.  
Each bin contains enough stars such that $\sum p_i = 50$.
The spatial error bars represent the radial extent of stars within each bin while the velocity dispersion errors are the 68\% confidence intervals. 
}
\label{fig:metallicity_keck}
\end{figure}

\begin{figure}
\includegraphics[width=\columnwidth]{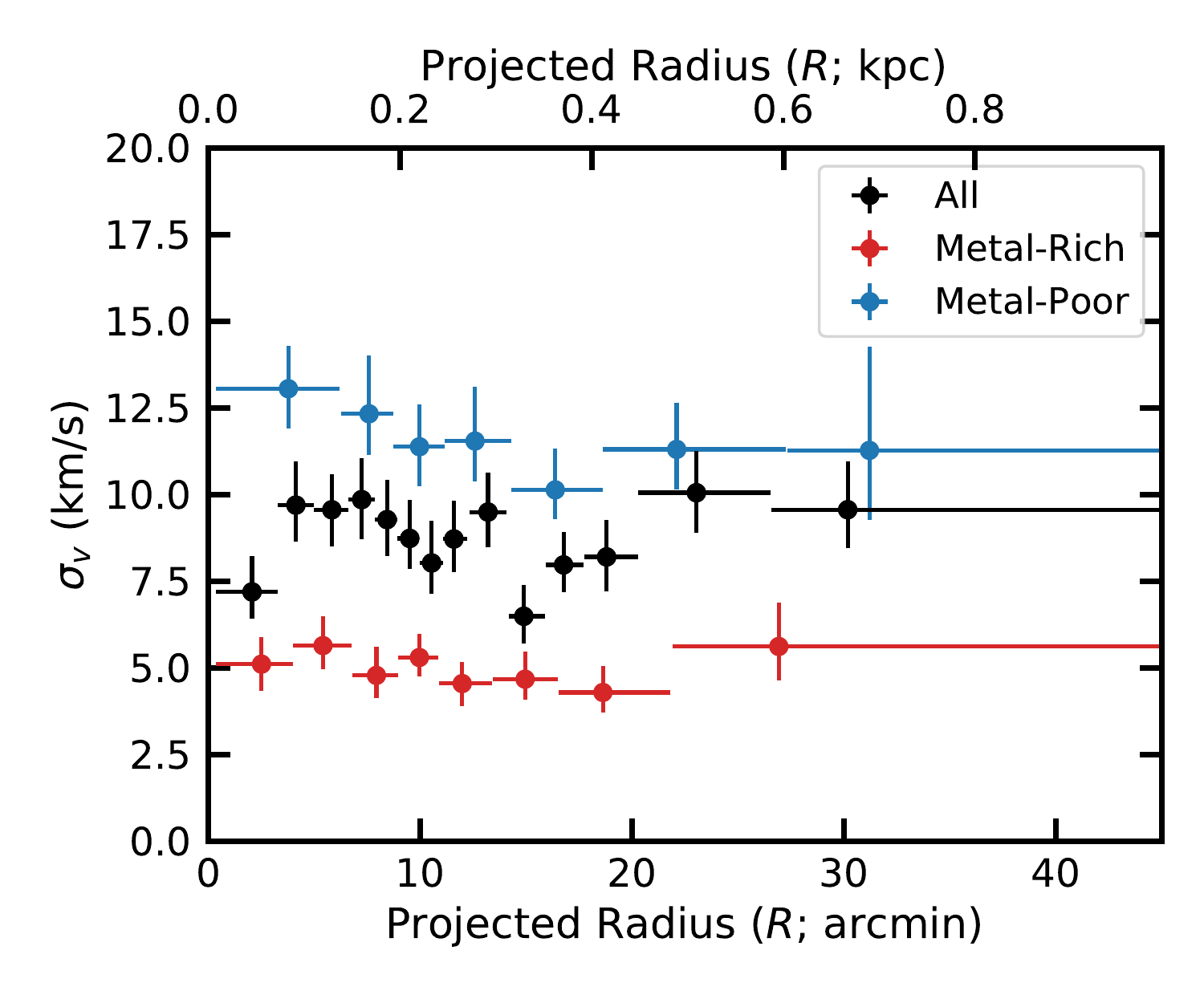}
\caption{Velocity dispersion ($\sigma_v$) in radial bins.
The average velocity is fixed to the best fit UMi value.  
The results are shown for the entire UMi sample (black)  and the metal-poor (blue) and metal-rich (red) populations.  
}
\label{fig:binned_velocity}
\end{figure}

In our  search for multiple chemodynamic populations, we  use RGB UMi members with metallicity measurements ($N=679$; $p_{\rm dSph}>0.95$).
Metallicity measurements are key for this analysis and horizontal branch stars may not have as robust metallicity measurements. 
To separate the populations, we set a prior between the velocity dispersions ($\sigma_{v, 1} < \sigma_{v, 2}$) and do not  assume any additional priors between the metallicity and/or spatial parameters.
We  assume there is no offset in mean line-of-sight velocity or  proper motion between the two populations. 
The posteriors of the  two population analysis are listed in Table~\ref{table:results} along with results from single component analysis with the same subset of RGB stars.

We find, with high significance ($\ln{B}=33.47$), two populations with distinct chemical, kinematic, and spatial distributions. 
The first population is kinematically cold ($\sigma_{v} = 4.9_{-1.0}^{+0.8} \kms$), more metal-rich ($\fehavr = -2.05\pm0.03$), and centrally concentrated, ($R_h = 221_{-17}^{+17} \, {\rm pc}$).
The second population is kinematically hot ($\sigma_{v} = 11.5_{-0.8}^{+0.9} \kms$), more metal-poor ($\fehavr = -2.29_{-0.06}^{+0.05}$) and spatially more extended ($R_h = 374_{-37}^{+49} \, {\rm pc}$).
The chemodynamical ordering of the two populations is the same as other dSphs  \citep[e.g., Sculptor;][]{Battaglia2008ApJ...681L..13B} despite only imposing a prior on $\sigma_v$. 
Although the metallicity separation is less than found in other dSphs, we will refer to the two stellar populations as the metal-rich and metal-poor populations. 
While the metallicity distributions overlap, the two stellar populations have distinct velocity, metallicity, and spatial distributions.

We show the spatial distribution of stars in each population in Figure~\ref{fig:ellipticity}.  
Interestingly, the metal-rich population is significantly more flattened ($\epsilon=0.75\pm0.03$) than the metal-poor population ($\epsilon=0.33_{-0.12}^{+0.09}$) and UMi in general ($\epsilon=0.55\pm0.01$). 
Both populations have  major axes of similar length.
In our standard analysis, we fixed the position angle.  We have explored  varying the position angle for each component and find that the difference in position angle between the two populations is small, $\Delta \theta < 5\degree$.  
The visual offset in Figure~\ref{fig:ellipticity} is due to the spectroscopic selection.
While we account for the spectroscopic selection function it is possible that there is a remaining bias. 
In Section~\ref{section:mmt} we examine and confirm the flattened metal-rich component with an independent UMi spectroscopic data set.

In Figure~\ref{fig:metallicity_keck}, we examine  the radial dependence of the average metallicity and metallicity dispersion of both populations.
The two populations are clearly offset in mean metallicity in all radial bins.
The difference in metallicity between the two populations is less than the differences observed in other dSphs (e.g., Fornax, Sculptor).
The metallicity dispersion is larger in the metal-poor population compared to the metal-rich population.

Figure~\ref{fig:binned_velocity} shows  the radial dependence of  $\sigma_v$ of the metal-rich and metal-poor populations.
The metal-rich $\sigma_v$ is constant with radius whereas the metal-poor $\sigma$ has hints that it deceases from  $\approx13\kms$ at the center to $\approx10\kms$ at large radii.  
For comparison, we also include the binned single-component velocity dispersion (black). 
Decreasing velocity dispersion profiles for sub-components are not unusual in a dSph.  They have been observed in the Fornax metal-poor population  \citep{Amorisco2012ApJ...756L...2A} and in the Sculptor metal-rich population   \citep{Battaglia2008ApJ...681L..13B}.

To test whether our assumption of a constant velocity dispersion affects the identification of stars within either chemodynamic population, we explore a model with a radial dependence. The velocity dispersion functional form we use is:  $\sigma_v(R_e)=\sigma_o \left(1 + R_e/R_{\sigma} \right)^{\alpha}$ (we use the same ellipticity for the spatial distribution and dispersion function).  This model has two additional parameters: a radial scale, $R_{\sigma}$, and a power law slope, $\alpha$.  The priors  for each parameter are $-2<\log_{10}{\left(R_{\sigma}/ 1\, {\rm kpc}\right)}<1$ and $-5<\alpha < 5$. 
Overall, the inferred  functional forms for the metal-rich and metal-poor are consistent with the binned profiles;  the metal-rich population is constant with radius whereas the metal-poor population decreases with radius from $\approx13\kms$ to $\approx10\kms$.
The posterior distribution for velocity dispersion parameters are degenerate with one another and non-kinematic parameters change little compared to the constant velocity dispersion model.
We find little change in the assignment of stars to either population.  The net absolute change ($\sum \Delta \lvert p_{\sigma (R_e)} - p_{\rm const}\rvert$) is $\sim24$ and the maximum absolute change of an individual star is small ($\Delta p_i=0.16$).  The mean and median differences of the membership are  both 0.03 and the standard deviation is 0.03.  Assuming a  constant velocity dispersion model does not  affect the identification of two populations or parameter inference.

Our UMi sample is built from three different observing epochs with  very different target selection criteria (a combination of different spatial regions and a different photometric input catalogs).  To explore whether the inhomogeneous    target selection is driving our inference of the chemodynamic populations we apply  the same analysis on three subsets.  Each subset excludes one of the three epochs and  the subset excluding the 2009, 2010, and 2012 data contains 435, 554, and 302 stars, respectively.  
Applying the same analysis to each subset, we continue to identify two chemodynamic populations with high significance ($\ln{\rm B}=23.57, 15.18, 25.14$). Our inhomogeneous target selection does not affect the inference of two chemodynamic populations.

\subsection{Analysis with MMT/Hechocelle Data}
\label{section:mmt}

\input{table_mmt.tex}

\begin{figure}
\includegraphics[width=\columnwidth]{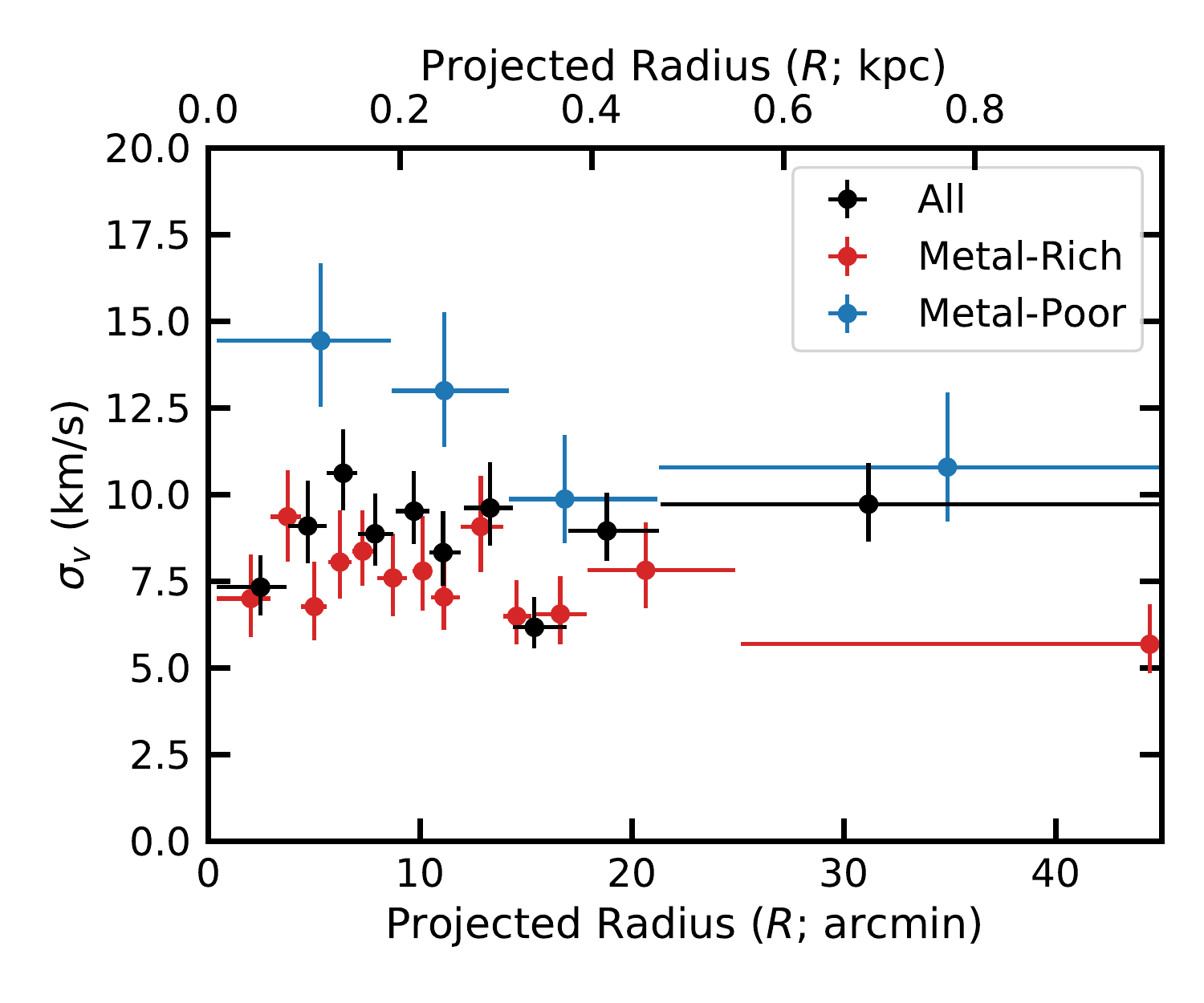}
\caption{Same as Figure~\ref{fig:binned_velocity} except with the MMT/Hectochelle  data set \citep{Spencer2018AJ....156..257S}.  The metal-poor components contain 25 stars per bin while the other components contain 40 stars.
}
\label{fig:binned_velocity_s18}
\end{figure}

As a cross check, we search for and find  two chemodynamic populations with the  independent UMi spectroscopic survey from MMT/Hectochelle observations  \citep{Spencer2018AJ....156..257S}.  
We first apply a similar mixture model to determine UMi membership and emphasize  again that   {\it Gaia} proper motions significantly improve  dSph and foreground separation.
The MMT sample contains brighter stars in general than the Keck sample and all but one star is in the  {\it Gaia} DR2 catalog.
We select candidate UMi stars with  a rough color-magnitude diagram selection with {\it Gaia} photometry ($G_{BP}-G_{RP}$ vs G) based on a slightly expanded selection in Figure 3 of \citet{GaiaHelmi2018A&A...616A..12G}.  
Using proper motion, velocity, metallicity, and spatial positions we identify 413 stars as UMi members  ($p_{\rm dSph}>0.95$).  Similar  to the Keck/DEIMOS data set, very few stars have intermediate membership; only 5 stars have a membership between $0.01<p_{\rm dSph}<0.95$.

The overall UMi properties\footnote{The mean velocity and metallicity we find for the MMT data are offset from the Keck value by $\approx2.3\kms$ and $\approx0.13$, respectively.  These are similar to the offsets found between common stars in Section~\ref{section:validation}.} we find in this data set are similar to the Keck/DEIMOS data set: $\overline{v}=-246.9\pm0.4\kms$, $\sigma_v=8.6\pm0.3\kms$, $\overline{[{\rm Fe/H}]}=-2.02\pm0.02$, $\sigma_{\rm [Fe/H]}=0.37\pm0.02$, $r_{p}=450_{-30}^{+33}\, {\rm pc}$, $\epsilon=0.56\pm0.03$, $R_{h}=297\pm15\, {\rm pc}$. 
We find a $\sigma_v$ value that is  $\approx0.6\kms$ larger than \citet{Spencer2018AJ....156..257S}.  This difference is likely due to  our use of {\it Gaia} DR2 astrometry and the different membership methods (mixture model versus  $\sigma$ clipping).

We find statistically significant evidence for two chemodynamic populations in the MMT/Hectochelle data set ($\ln{B}=17.94$).
The overall results are similar to the Keck/DEIMOS sample; the first population is  centrally-concentrated, dynamically-cold, and metal-rich whereas the secondary component is more extended, kinematically-hot, and metal-poor.
The properties of the two populations are summarized in  the last two columns of Table~\ref{table:results}.
In table~\ref{table:mmt}, we list our dSph and metal-rich membership for stars in the MMT sample.  
With the MMT data, we confirm the metal-rich population is aspherical ($\epsilon=0.64\pm0.04$) while the extended population is more spherical ($\epsilon=0.21_{-0.14}^{+0.16}$, constrained to be $\epsilon<0.38$ at the 90\% confidence interval).
We note that the two populations are closer in velocity dispersion ($\Delta\sigma_v \sim 4.4 \kms$) than our results in the Keck data set ($\Delta \sigma_v\sim 6.6 \kms$).
The MMT metal-poor $\sigma_v$ is consistent within errors with the Keck measurement whereas the metal-rich $\sigma_v$ is larger and disagrees at $\sim2\sigma$.
The metal-rich component is more metal-rich than the Keck results; this may be due to  offset found between the samples based on repeated measurements (see Section~\ref{section:validation} and Figure~\ref{fig:repeats_compare}). 
The MMT metal-poor population is more extended  but  more uncertain compared to the Keck results (due to the overall lower number of stars). 
We find an overall smaller fraction of stars in the metal-poor population ($f=0.24$ versus $f=0.46$).
The differences between the inferred properties may be based on differences in the target selection in the two data sets.

In Figure~\ref{fig:binned_velocity_s18}, we show the binned $\sigma_v$ profiles with the MMT data.  
The metal-poor $\sigma_v$  declines from  $\approx15\kms$ to $\approx 10 \kms$ and the metal-rich population is constant with radius.
This confirms the $\sigma_v$ profile of the metal-poor population that was seen in the Keck data.

The results from the MMT sample independently  confirm  the   two chemodynamical populations in  UMi.  
Moreover, it confirms two interesting features in UMi: a flattened metal-rich population with an almost spherical metal-poor component and a  declining $\sigma_v$ profile for the metal-poor population.
For the following analysis, we will analyze both UMi data sets independently.
We opt not to combine the data sets due to the observed offsets in velocity and metallicity  and the different methodologies for velocity and metallicity measurements.

\subsection{Search for Rotation}
\label{sec:rotation}

\begin{figure}
\includegraphics[width=\columnwidth]{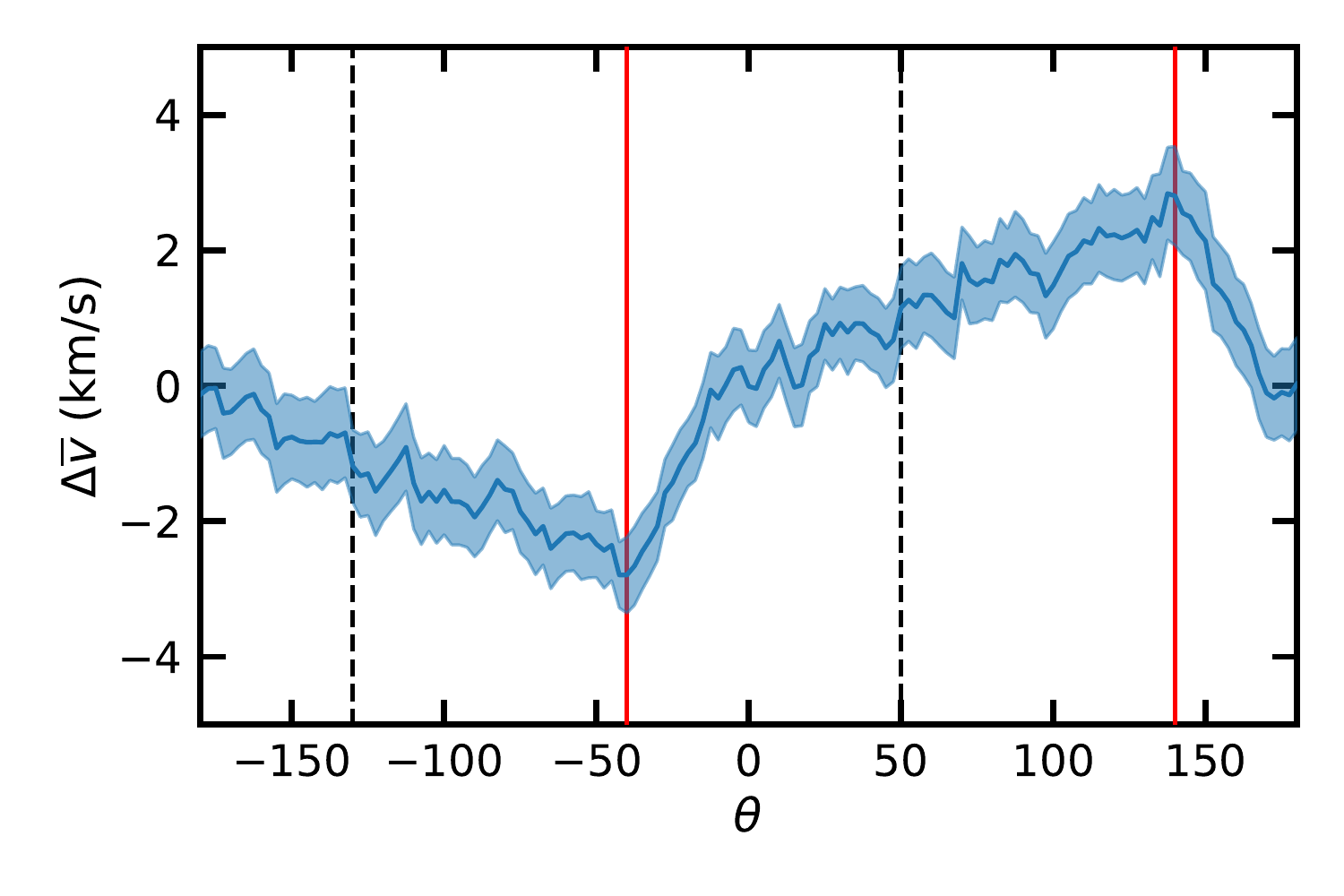}
\caption{Difference between the mean average velocity for each half  of the sample based on  bisecting lines for  different rotation axes for the MMT/Hectochelle data set. The angles $\theta=-130,50$ corresponds to the photometric major axis (dashed black lines) and the angles $\theta=-40, 140$ correspond to the photometric minor axis (solid red lines).  There are hints of rotation along the minor axis (i.e, prolate rotation).
}
\label{fig:rotation}
\end{figure}

The flattened nature of the metal-rich population is reminiscent of a disk galaxy and a natural question is whether it is rotating.
We first search for rotation with a simple test, splitting the sample in half based on  bisecting lines at different position angles and computing the difference between the mean velocity in each half. The entire Keck {\rm RGB} data set and the Keck metal-rich population show little to no signal whereas the Keck metal-poor population has an amplitude of $\sim2\kms$.  
In contrast, we observe a signal in the entire MMT data set, the MMT metal-rich population, and the MMT metal-poor population with amplitudes of $\sim3$, $\sim6$, and $\sim2 \kms$, respectively. In all cases with a signal, including the Keck metal-poor population,  the peak amplitude occurs near the photometric minor axis corresponding to prolate rotation ($\theta=-40, 140\degree$ corresponds to the photometric minor axis)\footnote{The \citet{Munoz2005ApJ...631L.137M} data set also shows evidence for rotation with an amplitude of $\sim5\kms$ that peaks at a similar position angle.}.  
We show the results of this test for the MMT data set in Figure~\ref{fig:rotation}.
We note that this is one of the better examples of rotation in the subsets examined.

To further quantify any potential rotation signal, we explore a rotation model that is constant with radius.  We add the rotation term to the relative velocity: $v_{\rm rel} = \overline{v} +  A_{\rm rot} \cos{\left(\theta - \theta_{\rm rot}\right)}$
\citep{Wheeler2017MNRAS.465.2420W}.  
This includes  two additional parameters, an amplitude, $A_{\rm rot}$, and a rotation angle, $\theta_{\rm rot}$.
In the Keck {\rm RGB} data set, the  rotation amplitudes we derive are consistent with zero and we find upper limits of $A_{\rm rot}  < 2.0 \,{\rm km s^{-1} kpc^{-1}}$, $A_{\rm rot} < 4.0\, {\rm km s^{-1} kpc^{-1}}$,  and $A_{\rm rot}  <2.2 \,{\rm km s^{-1} kpc^{-1}}$ (upper limits are at 95\% confidence) for the metal-rich population, metal-poor population, and the entire sample, respectively\footnote{We have explored this test by adding rotation to  one component at a time and both  components simultaneously  with similar results.}. 
In the MMT data set, we measure rotation amplitudes and rotation axes of $A_{\rm rot} = 1.9\pm0.9 \,{\rm km s^{-1} kpc^{-1}}$ and $\theta_{\rm rot}=156_{-17}^{+24} \,{\rm deg}$, $A_{\rm rot} = 3.6\pm2.0 \,{\rm km s^{-1} kpc^{-1}}$ and $\theta_{\rm rot}=164_{-33}^{+35} \,{\rm deg}$,  and $A_{\rm rot} = 1.8\pm0.8 \,{\rm km s^{-1} kpc^{-1}}$ and $\theta_{\rm rot}=156_{-19}^{+24} \,{\rm deg}$ for the metal-rich population, metal-poor population, and the entire sample, respectively.  
The rotation angle agrees with  the simple rotation test and is suggestive of prolate rotation.

We set upper-limits with the Keck data set (i.e., posterior of $A_{\rm rot}$ in the Keck data set is generally maximized at zero) and infer rotation with the MMT data set (i.e., the $A_{\rm rot}$ posterior peaks at a non-zero value). 
The different conclusion could be due to the larger spatial extent of the MMT data set and/or due to the differences in the velocity errors between the two instruments.
The median velocity error of the Keck and MMT data sets are $2.6\kms$ and $0.7\kms$, respectively.
Given these differences it is more likely that low levels of rotation could be observed with the MMT data set despite the larger Keck data set.

Interestingly, the rotation axis inferred from the MMT data is similar to  the velocity gradient induced by the perspective motion.  Based on the  proper motion measured from {\it Gaia} DR2, the maximum magnitude of perspective motion effect  is $\pm 0.2 \kms$.  This effect is already included in all of our modeling and  is much smaller than the rotation amplitudes we infer (see Section~\ref{section:likelihood}).  
Rotation is not favored in any of the data sets examined compared to the non-rotating model.
Additional extended data sets, especially along the minor axis, are required to confirm or refute the  hint of prolate rotation.

Prolate rotation has been previously observed in two  dwarf galaxies in the local: the M31 satellite dSph,  And~II \citep{Ho2012ApJ...758..124H} and the isolated transition type dwarf galaxy, Phoenix \citep{Kacharov2017MNRAS.466.2006K}.
In both cases, prolate rotation has been argued to be evidence for a recent merger.
For example, the peculiar kinematics of And~II may be due to a merger at $z\sim1.75$ \citep{Amorisco2014Natur.507..335A, Lokas2014MNRAS.445L...6L, delPino2017MNRAS.469.4999D, Fouquet2017MNRAS.464.2717F}.
In Phoenix, the spatial distribution of the young stars  are aligned  with the rotation axis/minor axis which is further evidence for a recent merger  \citep{Kacharov2017MNRAS.466.2006K}.
Prolate rotation has been observed for a subset of galaxies in the Illustris hydrodynamic simulation and the prolate rotation generally emerges from  late mergers with radial orbits \citep{Ebrova2017ApJ...850..144E}.
The hints of prolate rotation in UMi may be evidence for a recent  merger. 

\subsection{Exploring Additional Populations}

A natural question is whether UMi contains additional chemodynamic populations, similar to Fornax \citep{Amorisco2012ApJ...756L...2A} and Carina \citep{Kordopatis2016MNRAS.457.1299K}.
With our formalism it is straightforward to extend our analysis to an additional component.
We explore three different priors to disentangle populations: $\sigma_{v,1} <\sigma_{v,2}<\sigma_{v,3}$, ${\rm [Fe/H]}_{1} <{\rm [Fe/H]}_{2} < {\rm [Fe/H]}_{3}$, and the two first two priors combined.
For the fractions parameters, we use $f_x$ and $f_y$ as free parameters with the prior range: $0< f_{x,y}<1$. The transformations from these parameters  to the population fraction parameters are: $f_1=1-f_x$, $f_2=f_x (1-f_y)$, $f_3=f_x f_y$.

With the Keck data,  we find weak to moderate evidence in favor of the three population model when compared to the two population model  ($\ln{B}=3.09, \,  3.33, \, 1.61$ for the velocity dispersion, metallicity, and combined prior, respectively).
The different priors do not affect the  posterior distributions of the velocity dispersion, metallicity, and half-light radii of the three components and we observe the expected  chemodynamic ordering in all three cases. 
We will refer to the three components here as `1', `2', and `3' and ordering them from highest to lowest metallicity.
With the velocity dispersion prior, for the first population we find: $\sigma_{v,1}=4.7_{-0.9}^{+0.7}\kms$,  $\overline{[{\rm Fe/H}]}_1 = -2.00_{-0.04}^{+0.04}$, $\sigma_{\rm [Fe/H], 1}=0.25_{-0.03}^{+0.03}$, $\epsilon_1=0.75_{-0.03}^{+0.03}$, and $R_{h, 1} =214_{-17}^{+17} \,{\rm pc}$.
For the second population we find: $\sigma_{v,2}=9.9_{-1.4}^{+1.1}\kms$,  $\overline{[{\rm Fe/H}]}_2 = -2.29_{-0.04}^{+0.05}$, $\sigma_{\rm [Fe/H], 2}=0.09_{-0.05}^{+0.34}$, $\epsilon_2=0.54_{-0.25}^{+0.13}$, and $R_{h, 2} =347_{-57}^{+73} \,{\rm pc}$.
For the third population we find: $\sigma_{v,3}=11.8_{-0.9}^{+1.0}\kms$,  $\overline{[{\rm Fe/H}]}_3 = -2.31_{-0.07}^{+0.07}$, $\sigma_{\rm [Fe/H], 3}=0.42_{-0.31}^{+0.06}$, $\epsilon_3=0.27_{-0.16}^{+0.20}$, and $R_{h, 3} =377_{-58}^{+71} \,{\rm pc}$.
The metallicity dispersion posterior is multimodal in the second and third populations with peaks at $\sigma_{\rm [Fe/H]}\sim0.10$ and $\sigma_{\rm [Fe/H]}\sim0.45$. 
The fraction of stars in each component is: $f_1=0.46_{-0.09}^{+0.08}$, $f_2=0.26_{-0.06}^{+0.07}$, and $f_3=0.27_{-0.07}^{+0.08}$. 
The most metal-rich component in both the two and three population modeling have similar properties.
The `2' and `3' populations are similar to a splitting of the original `metal-poor' population.
Comparing with star-by-star membership confirms this picture, stars originally in the `metal-rich' component are more likely to be in population `1' while `metal-poor' stars are more likely to be in populations `2' or `3'.

With the MMT data, we find inconclusive to moderate evidence in favor of the three population model compared to the two population model ($\ln{B}=1.65, \,  2.68, \, 0.93$ for the sigma, metallicity, and combined prior, respectively).
In contrast to the Keck data, we find inconsistent results with the different priors analyzing the MMT data. 
Chemodynamic ordering is not observed between the three components with any of priors and is different depending on the prior used.
With all three priors the most metal-poor component (`3') has similar properties and  always has a larger spatial scale and velocity dispersion.
With the metallicity prior ($\overline{[{\rm Fe/H}]}_1<\overline{[{\rm Fe/H}]}_3<\overline{[{\rm Fe/H}]}_3$), we find: $\sigma_{v,3} > \sigma_{v,1} > \sigma_{v,2}$ and $R_{h,3} > R_{h,2} > R_{h,1}$.
With the velocity dispersion prior ($\sigma_{v,3} > \sigma_{v,2} > \sigma_{v,1}$), we find: $\overline{[{\rm Fe/H}]}_1\approx\overline{[{\rm Fe/H}]}_2<\overline{[{\rm Fe/H}]}_3$ and  $R_{h,3} > R_{h,1} > R_{h,2}$.
With the combined prior we find: $R_{h,3} > R_{h,1} \approx R_{h,2}$. 
The three population modeling with different priors does not have  consistent results with the MMT data.

Our three-component results for the Keck data set are consistent across the three different priors we used and the Bayesian evidence favors three components moderately. This moderate evidence must be interpreted with care. Even if more data would increase the evidence for three components, it would not necessarily argue for three distinct chemodynamic populations. The bimodal posterior of $\sigma_{\rm [Fe/H]}$ seems likely to be driven by the non-Gaussian distribution of ${\rm [Fe/H]}$ for the metal poor population.
Chemical evolution models are known to produce non-Gaussian metallicity distributions \citep[e.g.][]{Kirby2011ApJ...727...78K}, hence the conservative interpretation is that we have two distinct chemodynamic populations with more complex metallicity distributions.
A straight-forward test is to examine a two population model where the metallicity distribution of the metal-poor component contains two Gaussians instead of one.
With this model we find that mean metallicities are equivalent ($\overline{\rm Fe/H}_{\rm MP, 1}=\overline{\rm Fe/H}_{\rm MP, 2} = -2.30_{-0.04}^{+0.06}$) and the metallicity dispersion parameter mirror the results with the three population model ($\sigma_{\rm [Fe/H], MP, 1}=0.09_{-0.04}^{+0.04}$ and $\sigma_{\rm [Fe/H], MP, 2}=0.48_{-0.04}^{+0.04}$).
This model is a better fit compared to the three component models ($\ln{B}=2.83$ compared to the best fitting three component model).
There is some evidence that the axis ratios could be different for the second and third components, but this could be an indication that the luminosity profile for the metal-poor component deviates from the assumed Plummer profile. As points of comparison, we note that the two dSphs with three reported chemodynamical populations, Carina and Fornax both have larger stellar masses and have more extended star formation histories than UMi \citep{Weisz2014ApJ...789..147W}. In addition, the reported chemodynamic populations have larger metallicity differences than our inferences for the second and third components. Given the reasons above, we conclude that the evidence for a distinct third chemodynamic population in UMi is weak.

\section{Simple  Estimators  of Mass Slope}
\label{section:slope}

\begin{figure}
\includegraphics[width=\columnwidth]{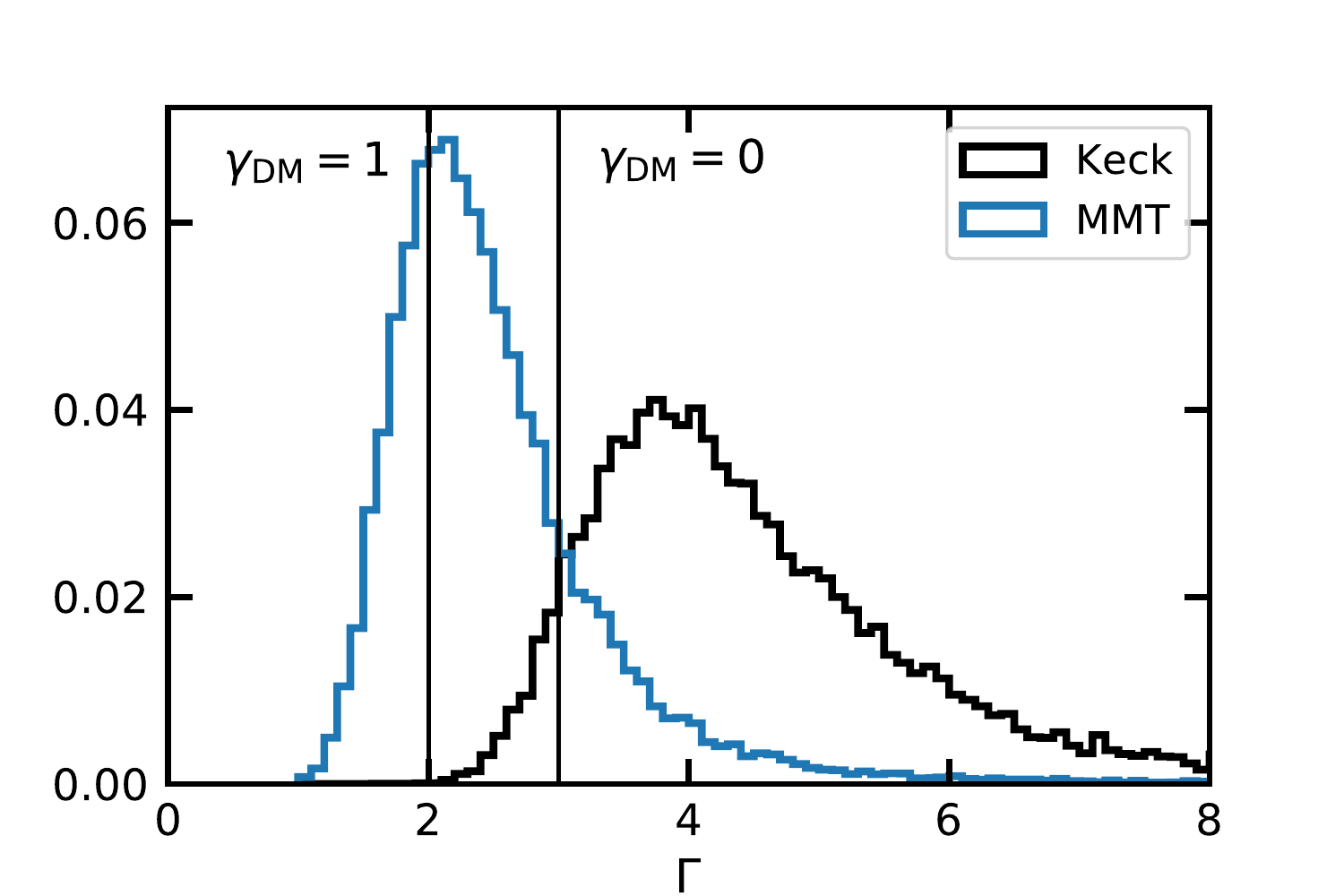}
\caption{Posterior distributions of the mass slope ($\Gamma$)  for the Keck/DEIMOS (black) and MMT/Hectochelle (blue) data sets.  
The two lines show  where `cuspy' (left; $\Gamma=2$) and `cored' (right; $\Gamma=3$) halos  lie in the $\Gamma$ distribution. 
The Keck data set favors very large $\Gamma$ which may indicate the break down of the simple estimator (i.e., non-constant velocity dispersion in the metal-poor component or the large ellipticity of the metal-rich component). 
The MMT data set favors a `cuspy' distribution but does not exclude a `cored' dark matter halo.
}
\label{fig:gamma}
\end{figure} 

Although the overall mass profile is degenerate with the stellar anisotropy, the mass within the stellar half-light radius, $M_h = M(r_h)$,  is well-determined for spherical systems \citep{Walker2009ApJ...704.1274W, Wolf2010MNRAS.406.1220W}.  
The difference in mass at the  half-light radii of each chemodynamic population provides an estimate for the mass slope:  $\Gamma = \Delta M_h / \Delta r_h$ \citep{Walker2011ApJ...742...20W}.  
To accommodate the assumption of spherical symmetry, we use the geometric mean of the major and minor axes as a ``spherical" half-light radius (azimuthally averaged half-light radius, $R_h = r_p \sqrt{q}$). 
The modified mass-slope estimator is:

\begin{equation}
\Gamma = 1 + \frac{\ln{\left[\sigma^2_{v,1}/\sigma^2_{v,2}\right]}}{\ln{\left[\left(r_{p,1} \sqrt{q_1}\right) /\left(r_{p,2}\sqrt{q_2}\right)\right]}}.
\label{eq:walker_mass_slope}
\end{equation}

\noindent  The benefit of this method is that no additional dynamical modeling is required; only quantities directly measured in the two component analysis are used.
However, this estimator may not be valid for UMi.  
The potential stellar rotation, extreme ellipticity, and non-constant $\sigma_v$ invalidate several of the estimator's assumptions.  
While we have accounted for the ellipticity, these estimators have not been tested at high ellipicity.
Note that $\Gamma$ is not a direct measurement of the inner density profile (at $r = 0$) but a measurement of the average slope of the mass profile between $R_h$ of the two populations.
For a dark matter density profile scaling as $\rho_{\rm DM} \propto r^{-\gamma_{\rm DM}}$ in the central regions, $\Gamma$ places an upper limit on the density-slope as $\gamma_{\rm DM} < 3 - \Gamma$ \citep{Walker2011ApJ...742...20W}.
For a NFW (or `cuspy') profile: $\gamma_{\rm DM}=1$ and $\Gamma=2$.
For a constant (`cored') density profile: $\gamma_{\rm DM}=0$, $\Gamma=3$.

With these caveats, we measure $\Gamma = 4.2_{-0.9}^{+1.6}$ in the radial range of $\approx200-400$ pc with the Keck data set. 
We display the posterior distributions of $\Gamma$ in Figure~\ref{fig:gamma}. 
This disfavors a `cuspy' NFW halo ($\Gamma=2$) at greater than $2\sigma$ but  the majority of the posterior has $\Gamma > 3$,  implying a unphysical density that increases with radius (assuming $\rho_{\rm DM} \propto r^{-\gamma_{\rm DM}}$). 
This tendency to favor large $\Gamma$ is due to the large difference in $\sigma_v$ but smaller relative difference in $R_h$.
These quantities could mis-estimated due to the non-constant $\sigma_v$ or the large ellipticity.
For comparison we find $\Gamma=2.3_{-0.5}^{+0.8}$ with the MMT data set, which favors a `cuspy' halo but it is not precise enough to distinguish between `cored' or `cuspy' halos.  
The $\Gamma$ posterior distributions of the Keck and MMT data sets overlap but favor different interpretations of the inner dark matter slope.
The disagreement between the two data sets is caused by the difference in metal-poor population $R_h$ and the difference in metal-rich population $\sigma_v$.  Simply increasing the Keck metal-poor $R_h$ by 50\%  results in much better $\Gamma$ agreement better the two data sets.  A more extended data set will be key to determining the inner slope of the dark matter density profile.

The half-light mass estimators that $\Gamma$ is based on have been verified to be robust  in a wide range of systems and N-body simulations  \citep{Walker2011ApJ...742...20W, Laporte2013MNRAS.433L..54L, Kowalczyk2013MNRAS.431.2796K, Lyskova2015MNRAS.450.3442L, Campbell2017MNRAS.469.2335C, Gonzalez-Samaniego2017MNRAS.472.4786G, Errani2018MNRAS.481.5073E}.
There has been a healthy discussion about the robustness of the $\Gamma$ estimator \citep{Walker2011ApJ...742...20W, Laporte2013MNRAS.433L..54L, Kowalczyk2013MNRAS.431.2796K, Genina2018MNRAS.474.1398G}.
More recent results on the viewing angle suggest that there may be biases in the estimator \citep{Walker2011ApJ...742...20W, Laporte2013MNRAS.433L..54L}.

\section{Chemodynamic Populations Throughout the Local Group} 
\label{section:chemo_compare}

Multiple distinct global chemodynamic populations\footnote{Localized kinematic or chemodynamical substructure has been seen or claimed in multiple dSphs. However, detections are not always consistent between different data sets and methods \citep{Kleyna2003ApJ...588L..21K, Kleyna2004MNRAS.354L..66K, Walker2006ApJ...642L..41W,Battaglia2011MNRAS.411.1013B, Fabrizio2011PASP..123..384F, Amorisco2014Natur.507..335A, Pace2014MNRAS.442.1718P, delPino2017MNRAS.469.4999D, delPino2017MNRAS.465.3708D, Cicuendez2018MNRAS.480..251C, Lora2019ApJ...878..152L, Kim2019ApJ...870L...8K}. 
} have been found with high significance in several dSphs. 
Chemodynamic populations are  structured such that the inner central concentrated population  is metal-rich and kinematically cold while the outer extended population is metal-poor and kinematically hot.
The two populations that we have uncovered in UMi follow these same trends.
Chemodynamical populations have encouraged a vast amount of dynamical modeling analysis \citep[e.g.,][]{Battaglia2008ApJ...681L..13B, Agnello2012ApJ...754L..39A, Amorisco2012MNRAS.419..184A, Strigari2017ApJ...838..123S, Hayashi2018MNRAS.481..250H}.

Multiple populations have  been detected in: Carina\footnote{The intermediate and metal-rich populations in Carina are an exception to the general chemodynamic ordering trend. The intermediate metallicity population is the most compact and kinematically cold \citep{Kordopatis2016MNRAS.457.1299K}.  It is unclear why in Carina the populations are mixed in this manner relative to other dSphs. 
The star formation in Carina is episodic and so distinct and well-separated that they can be clearly seen without any special modeling \citep{Hurley-Keller1998,deBoer2014A&A...572A..10D}. 
Furthermore,  \citet{Walker2011ApJ...742...20W} have a null detection in Carina.} \citep{Kordopatis2016MNRAS.457.1299K, Fabrizio2016ApJ...830..126F, Hayashi2018MNRAS.481..250H}, Fornax \citep{Battaglia2006A&A...459..423B, Walker2011ApJ...742...20W, Amorisco2012ApJ...756L...2A, Amorisco2013MNRAS.429L..89A, Hendricks2014A&A...572A..82H}, Sculptor \citep{Tolstoy2004ApJ...617L.119T, Battaglia2008ApJ...681L..13B, Walker2011ApJ...742...20W,  Zhu2016MNRAS.463.1117Z}, Sagittarius\footnote{Chemodynamic trends  have also been observed in the leading and trailing arms of the  Sagittarius stellar stream \citep{Gibbons2017MNRAS.464..794G}. } \citep{Majewski2013ApJ...777L..13M} and Ursa Minor (this work).  
There are claims or lower significance detections in several other dSphs.
For example, \citet{Ibata2006MNRAS.373L..70I} claim a detection in Canes Venatici~I. However, larger data  sets do not confirm this feature \citep{Simon2007ApJ...670..313S, Ural2010MNRAS.402.1357U}.  
Similarly, \citet{Koposov2011ApJ...736..146K} find that Bo\"{o}tes~I favors a two-component model in kinematics while previous data sets did not find this \citep{Munoz2006ApJ...650L..51M, Martin2007MNRAS.380..281M}. 
There is tentative evidence in the local field isolated dSphs Cetus \citep{Taibi2018A&A...618A.122T} and Tucana \citep{Taibi2020arXiv200111410T} and  hints in Leo~II \citep{Spencer2017AJ....153..254S} and Sextans \citep{Battaglia2011MNRAS.411.1013B} that may be comfirmed with larger sample sizes.

While the chemodynamical populations that we have uncovered in UMi are similar to the standard chemodynamic ordering, there are some interesting differences. 
First, the metallicity distributions in UMi are closer than in other dSphs.  
This could be due to a merger of two  similar sized galaxies or there was little gas enrichment before the formation of the second component.
Second, the UMi metal-rich population is flattened and significantly more  elongated than the more spherical metal-poor population.  
UMi is more elongated ($\epsilon=0.56$) than the other classical ($0.07<\epsilon<0.45$) satellites   \citep{Munoz2018ApJ...860...66M}.
Chemodynamic analyses of other dSphs have found similar ellipticities for  each  population  \citep[Carina, Fornax, Sculptor;][]{Zhu2016MNRAS.463.1117Z,Kordopatis2016MNRAS.457.1299K,  Hayashi2018MNRAS.481..250H}. 
Third, the metal-poor $\sigma_v$ decreases $\approx5\kms$ to $\approx 10 \kms$ beyond $\approx300 \,{\rm pc}$.
Declining $\sigma_v$ have been observed in the Fornax metal-poor population  \citep[][]{Amorisco2012ApJ...756L...2A} and the Sculptor metal-rich population   \citep{Battaglia2008ApJ...681L..13B}  chemodynamic analysis.
It is interesting that as a whole the classical dSphs have constant $\sigma_v$ with radius \citep{Walker2007ApJ...667L..53W} whereas some sub-populations are observed to deviate from this trend.
Each of these characteristics may be related to the formation mechanism of the two populations.

Several formation scenarios for multiple populations have been proposed including: mergers \citep{Amorisco2012ApJ...756L...2A, delPino2015MNRAS.454.3996D, Genina2019MNRAS.488.2312G}, supernova feedback \citep{Salvadori2008MNRAS.386..348S, Revaz2009A&A...501..189R},  tidal interactions \citep{Pasetto2011A&A...525A..99P}, interactions with a gaseous cosmic filament \citep{Genina2019MNRAS.488.2312G}, or compression of gas   during infall  \citep{Genina2019MNRAS.488.2312G}.
In addition, they are seen in  isolated hydrodynamical simulations \citep{Revaz2018A&A...616A..96R}.
In the merger scenario, the spatial separation is due to  metal-poor stars migrating to outer orbits with the metal-rich population forming in-situ after the merger \citep{Genina2019MNRAS.488.2312G}.
Based on {\it Gaia} DR2 proper motions, the infall time of UMi is $10.7$ Gyr \citep{Fillingham2019arXiv190604180F} and is early enough that the metal-rich population could have formed due to gas compression from  ram pressure stripping during MW infall.
Large differences between the shapes of two populations are not seen in simulated dSph satellites \citep{Genina2018MNRAS.474.1398G}.
In field dwarfs, the metal-poor populations are more spherical than the metal-rich populations. However, this difference is smaller in satellites, likely due to tidal stripping \citep{Genina2018MNRAS.474.1398G, Genina2019MNRAS.488.2312G}.
At the lowest stellar masses, \citet{Genina2019MNRAS.488.2312G}, find that most multiple populations in dSphs are primarily formed due to mergers.  
Observations of prolate rotation may be evidence for  past mergers \citep{Amorisco2014Natur.507..335A, Kacharov2017MNRAS.466.2006K} and the two populations in  UMi could be evidence for a late time merger.
We note that \citet{Genina2019MNRAS.488.2312G} conclude that spatial and kinematic information is insufficient to determine the  formation mechanism but metallicities and star-formation histories can provide clues on their origin.  
The multiple populations in UMi could have formed due to mergers or gas interactions but it is not yet conclusive which mechanism is responsible.  
Future star formation history constraints combined with  metallicity distribution functions will assist in  disambiguating between these scenarios.

\section{Dark Matter Annihilation and Decay Rates of Ursa Minor }
\label{section:j_factor}

\begin{figure}
\includegraphics[width=\columnwidth]{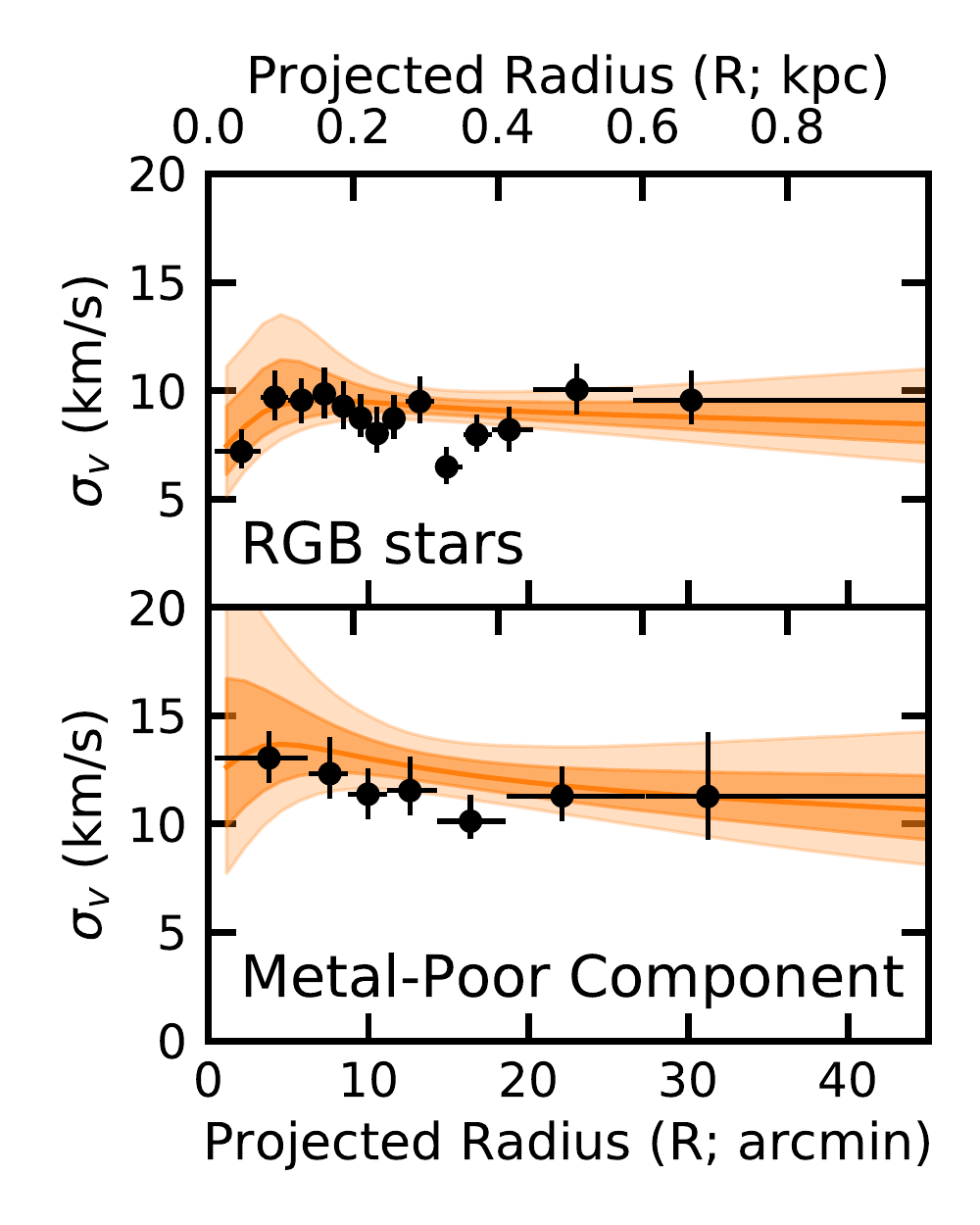}
\caption{ Binned velocity dispersion profiles (black points) overlaid with spherical Jeans  fits (orange lines/bands) for the Keck/DEIMOS data set.   
The shaded bands show the $1-\sigma$ and $2-\sigma$ confidence intervals.  
The top (bottom) panel shows the RGB (metal-poor) stars.
Note that the binned velocity dispersion profiles are for visualization only and the fit is done on a star-by-star basis. 
}
\label{fig:jeans_results_keck}
\end{figure}

\input{table_jfac.tex}

The close proximity and high dark matter content of the MW dSphs make them ideal candidates to search   for signatures   of dark matter annihilation or decay \citep{Baltz2000, Tyler2002, Evans2004PhRvD..69l3501E, Bergstrom2006, Colafrancesco2007, Strigari2007PhRvD..75h3526S}.  
The dark matter annihilation or decay flux from a dSph depends on the distribution of dark matter within the dSph (the astrophysical component) and properties of the dark matter particle such as the mass or cross-section (the particle physics component). 
For velocity independent  models these components are separable and the astrophysics portion is referred to as the J- or D-factor for annihilation and decay, respectively. 
The J- and D-factors are:

\begin{equation}
J(\theta_{\mathrm{max}}) = \underset{\mathrm{los}}{\iint} \rho_{\mathrm{DM}}^2 (l, \Omega) \, \mathrm{d}l \mathrm{d}\Omega\,,
\label{eq:j_fac}
\end{equation}

and 

\begin{equation}
D(\theta_{\mathrm{max}}) = \underset{\mathrm{los}}{\iint} \rho_{\mathrm{DM}} (l, \Omega) \, \mathrm{d}l \mathrm{d}\Omega\,.
\end{equation}

\noindent The integrals are over the line-of-sight direction and a solid angle centered on the dwarf.  $\theta_{\mathrm{max}}$ is the maximum angle probed.  
The standard methodology for determining $\rho_{\rm DM}$ for dSphs is with dynamical models based the spherical Jeans equation \citep[e.g.,][]{Strigari2007PhRvD..75h3526S, Walker2011ApJ...733L..46W, Bonnivard2015MNRAS.453..849B, Geringer-Sameth2015ApJ...801...74G, Pace2019MNRAS.482.3480P}.
For a spherically symmetric system in dynamical equilibrium the spherical Jeans equations reduce to a single differential equation \citep{BT2008}:

\begin{equation}
\frac{\mathrm{d}\nu \sigma_r^2}{\mathrm{d}r} + \frac{2}{r} \beta(r) \nu \sigma_r^2 + \nu \frac{G M(r)}{r^2} = 0\,,
\label{eq:jeans}
\end{equation}

\noindent where $\nu(r)$ is the 3D stellar density distribution, $M(r)$ is the mass distribution, $\sigma_r$ is the radial velocity dispersion, and   $\beta$ is the stellar anisotropy, which is a function of the tangential and radial velocity dispersion, $\beta(r)=1 -\sigma_t^2(r)/\sigma_r^2(r)$.
To compare the theoretical velocity dispersion profiles to observed line-of-sight velocity data, we project the radial dispersion  into the line-of-sight direction,

\begin{equation}
\Sigma(R) \sigma_{\mathrm{los}}^2 (R) = 2 \int_R^{\infty}\mathrm{d}r \, \left(1 - \beta(r) \frac{R^2}{r^2} \right) \frac{r \, \nu \sigma_r^2(r)}{\sqrt{r^2 - R^2}}\,,
\end{equation}

\noindent where $R$ is the projected radial distance, $\Sigma(R)$ is the projected tracer distribution (2D), and 
$\sigma_{\mathrm{los}}^2$ is the velocity dispersion in the line-of-sight direction.  

We assume a Plummer profile for the tracer profile \citep{Plummer1911MNRAS..71..460P}. The 3-D spherically symmetric analog to the projected (2D) Plummer profile (Equation~\ref{eq:plummer}) is:

\begin{equation}
\nu (r) = \frac{3}{4 \pi r_{\rm p}^3} \left[ 1 + \left( r/r_{\rm p} \right)^2\right]^{-5/2}\,.
\end{equation}

\noindent  To approximate spherical symmetry we use the spherically averaged half-light radii ($R_h = r_p \sqrt{q}$).
We assume that the mass profile is entirely dominated by the dark matter halo.  We assume an Einasto dark matter halo profile  \citep{Einasto1965TrAlm...5...87E,Navarro2004MNRAS.349.1039N},
\begin{equation}
\rho_{\mathrm{Einasto}}(r) = \rho_{-2} \exp{\left\{-\frac{2}{\alpha} \left[ \left(\frac{r}{r_{-2}} \right)^{\alpha}  - 1 \right] \right\} } \,,
\end{equation}
\noindent where $r_{-2}$, $\rho_{-2}$, and $\alpha$ are the scale radius, scale density, and  slope parameter, respectively.
For the stellar anisotropy profile we use the generalized \citet{Baes2007A&A...471..419B} profile,

\begin{equation}
\beta(r) = \frac{\beta_{0} + \beta_{\infty} (r/r_{\beta})^{\eta}}{1 + (r/r_{\beta})^{\eta}} \,,
\label{equation:beta}
\end{equation}

\noindent  where $\beta_0$, $\beta_{\infty}$, $r_{\beta}$, and $\eta$ are  parameters that  correspond to the inner anisotropy, outer anisotropy, radial anisotropy scale, and transition slope, respectively.

We assume an unbinned likelihood \citep{Strigari2008ApJ...678..614S, Geringer-Sameth2015ApJ...801...74G}.
We weight each star by its membership probability (Equation~\ref{eq:member}).

To set the maximum size of the  dark matter halo (for Equation~\ref{eq:j_fac}), we calculate the satellite's tidal radius,
$r_t= d \left( M_{\rm dSph}/M_{\rm MW}(d) (2 - \frac{{\rm d}\ln{M_{\rm MW}}}{{\rm d}\ln r} ) \right)^{1/3}$ \citep{Springel2008MNRAS.391.1685S}.
We use the MW total mass profile from \citet{Eadie2016ApJ...829..108E}. 
We find $r_t\approx 5-6$ kpc for UMi.
The majority of the J-factor comes from the inner region of the dark matter halo and small changes in  $r_t$ do not significantly impact J.
We note that the UMi has one of the lowest pericenters of the classical satellites ($r_{\rm peri}\approx30$ kpc), and a more accurate $r_t$ calculation should account for the full orbit of UMi \citep{GaiaHelmi2018A&A...616A..12G, Fritz2018A&A...619A.103F}.

We examine two samples for each data set, the entire RGB sample and the metal-poor population.  
We do not dynamically model the metal-rich population or both populations simultaneously  due to the large ellipticity in the metal-rich component.  
The metal-poor population is much more spherical and spherical Jeans may have some unknown systematics when modeling highly flattened systems. 
We model the RGB sample to provide a comparison sample that is selected in a similar manner to literature UMi J-factor measurements.
Our spherical Jeans fits for the Keck data samples are shown in Figure~\ref{fig:jeans_results_keck} and can adequately explain the stellar kinematics.
For the J-factor ($\log_{10}{J(\theta=0.5\degree)}=\log_{10}{J}$), we derive  $\log_{10}{J}=18.73_{-0.11}^{+0.17}$ and $\log_{10}{J}=18.47_{-0.09}^{+0.13}$ for the RGB samples for the Keck and MMT data sets, respectively.
Our results for the metal-poor component are larger.  
We derive $\log_{10}{J}=19.08_{-0.13}^{+0.16}$ and $\log_{10}{J}=18.61_{-0.31}^{+0.31}$ for the metal-poor populations for the Keck and MMT data sets, respectively.
Our J- and D-factor results for additional angles are summarized in Table~\ref{table:j_fac}.
Several recent literature measurements of the J-factor are: $\log_{10}{J}=18.93_{-0.19}^{+0.27}$  \citep{Geringer-Sameth2015ApJ...801...74G}, $\log_{10}{J}=18.8 \pm 0.19$ \citep{FermiLATCollaboration2014PhRvD..89d2001A}, $\log_{10}{J}=19.0 \pm 0.1$ \citep{Bonnivard2015MNRAS.453..849B}, $\log_{10}{J}=18.75 \pm 0.12$
\citep{Pace2019MNRAS.482.3480P}, and  $\log_{10}{J}=18.75_{-0.13}^{+0.17}$  \citep{Horigome2020arXiv200204866H}.
In general, our Keck results agree with recent values in the literature however our MMT results are smaller.

The J-factor has a simple scaling based on the velocity dispersion, distance, and stellar size: $J \propto \sigma_v^4/d^2/r_{1/2}$ \citep[][]{Pace2019MNRAS.482.3480P}\footnote{See   \citet{Evans2016PhRvD..93j3512E}, and \citet{Ullio2016JCAP...07..025U} for similar discussion. }.
The differences in these inputs  can explain a large portion of these differences.  
Generally, our assumed distance is further away, the stellar size larger, and the velocity dispersion smaller than other studies which will all  decrease our measurement compared to the literature measurements.
As a direct comparison we have repeated the analysis in \citet{Pace2019MNRAS.482.3480P} for the Keck and MMT data set. \citet{Pace2019MNRAS.482.3480P} assumed an NFW dark matter density profile, a constant stellar anisotropy, used the stellar parameters from \citet{Munoz2018ApJ...860...66M}, and varied the distance with a Gaussian prior. 
We derive $\log_{10}{J}=18.58\pm0.12$ and $\log_{10}{J}=18.64\pm0.11$ with the Keck and MMT data sets, respectively, with these modeling assumptions compared to $\log_{10}{J}=18.75 \pm0.12$.
The UMi data\footnote{This was a  preliminary catalog of MMT/Hectochelle data eventually published in \citet{Spencer2018AJ....156..257S}.} utilized by \citet{Pace2019MNRAS.482.3480P}  contained 311 stars with $\sigma_v=9.3\pm0.4\kms$ compared to $\sigma_v=8.6\kms$ in the Keck and MMT data sets.  The decrease in J-factor due to the lower velocity dispersion is expected to be $\Delta \log_{10}{J} \sim -0.18$ or $\log_{10}{J} \sim 18.63$, which agrees with the J-factor calculation.
Compared to the other data sets, we estimate differences of $\Delta \log_{10}{J}\sim0.43, 0.31$ just due to the changes in these inputs in  \citet{Bonnivard2015MNRAS.453..849B}, \citet{Geringer-Sameth2015ApJ...801...74G}, respectively.
Furthermore, we note that our structural parameters are derived from the spectroscopic sample alone and are more uncertain than those derived from photometric catalogs due to the lower number of stars, improving the constraints on the metal-poor stellar distribution is a straightforward manner to improve the J-factor measurement. 
Other modeling assumptions such as the dark matter halo, velocity anisotropy, or stellar distribution can account for the remaining differences.

The J-factors derived from the metal-poor component are larger compared to the sample as a whole (RGB data set).  Although the metal-poor component is more extended than UMi as a whole the larger  $\sigma_v$ implies an overall denser dark matter halo.  The spherical Jeans modeling is  more applicable to the more  spherical metal-poor population than UMi as a whole because of the ellipticity.
This suggests that future work with axisymmetric mass models will be particularly beneficial  for understanding UMi \citep[e.g.,][]{Zhu2016MNRAS.463.1117Z, Klop2017PhRvD..95l3012K, Hayashi2018MNRAS.481..250H}. 
Of the classical dSphs, UMi has one of the largest J-factors, our results suggest that it continues to be an excellent target, however, its kinematics are  more complex than previously thought.

\section{Conclusion}
\label{sec:conclusion}

We have presented  line-of-sight velocities and stellar metallicities from the largest  spectroscopic data set of the classical dSph Ursa Minor. 
Our Keck/DEIMOS observations include 1630 measurements of the line-of-sight velocity  and 1389 metallicity measurements  of 1462 unique stars.
Through  a dSph and MW foreground mixture model, we utilized a combination of velocity, metallicity, position, and proper motion  to  identify 892 UMi members,  doubling the number of known spectroscopic members.

We have discussed a methodology for  disentangling chemodynamical populations in dSph galaxies  building upon previous work by extending the analysis to axisymmetry and utilizing Bayesian evidence to compare models with and without multiple populations. 
We have uncovered two chemodynamic stellar populations in UMi at high statistical significance ($\ln{B} =  33.47$).
The first population is more metal-rich ($\fehavr \approx -2.05$), kinematically cold  ($\sigma_v \approx4.9\kms$), and centrally concentrated  ($R_h\approx220$ pc)  compared to the  second more metal-poor ($\fehavr \approx -2.29$), kinematically hot ($\sigma_v \approx11.5\kms$), and extended population ($R_h\approx370$ pc).
The two  populations in UMi follow the same chemodynamical ordering observed in other dSphs \citep[e.g., Sculptor and Fornax;][]{Battaglia2008ApJ...681L..13B, Walker2011ApJ...742...20W}.

We applied the same methodology to a smaller independent  MMT/Hectochelle spectroscopic data set  \citep{Spencer2018AJ....156..257S}.  
Our analysis of this sample confirmed our discovery of two chemodynamic populations in UMi ($\ln{\rm B} = 17.94$).  
The properties of the chemodynamical  populations with the MMT data set are overall similar to the Keck/DEIMOS results although  we find that the metal-rich velocity dispersion and the metal-poor spatial scale  are larger compared to the Keck data set. 

In both data sets, the UMi metal-rich population is significantly more  elongated ($\epsilon=0.75\pm0.03$) than the almost spherical metal-poor population.  The large difference in ellipticity may be a hint of different  formation mechanisms. 
The velocity dispersion of the metal-poor population decreases from $\approx13\kms$ at the center of UMi to $\approx10\kms$ at the edge of our sample.
We searched for and found some hints of prolate rotation in UMi in the MMT data set.  However, in the Keck data set we found no evidence for rotation.
We further explored whether a three population chemodynamic model better explains the UMi data. 
There is some evidence of three populations in the Keck data set however the `new' population is a split of the original metal-poor component suggesting that the metal-poor component may have a non-Gaussian metallicity distribution. In contrast, the three population results with the MMT data do not have consistent results with different priors and disagree with the Keck results. 
Additional extended data (especially along the minor axis) will be key for determining if  UMi exhibits prolate rotation  or if UMi is split into additional chemodynamic populations.  
Future spectroscopic observations can utilize {\it Gaia} astrometry to remove a large portion of the MW foreground and increase the yield at large radii.

We  explored  a simple mass-slope estimator, $\Gamma$ \citep{Walker2011ApJ...742...20W}, to probe the dark matter distribution.
The $\Gamma$ distribution disfavors a `cuspy' halo with the Keck data. However, the naive interpretation of very large $\Gamma$ values implies that the density increases with radius. 
In contrast, the MMT data favors a `cuspy' slope but is still consistent with `cored' halos.
The flattened metal-rich population, potential stellar rotation,  and the non-constant metal-poor velocity dispersion invalidates several of the assumptions  this estimator is based on. More detailed modeling is required to robustly determine the dark matter density distribution in UMi.
In addition, it would be interesting to combine line-of-sight velocities with proper motion measurements obtained from {\it Gaia} to better constrain the inner slope of the dark matter mass profile mass slope \citep{Strigari2007ApJ...657L...1S, Massari2018NatAs...2..156M, Massari2020A&A...633A..36M}.

We have calculated the astrophysical components (J and D-factors) for dark matter annihilation or decay based on the inferred dark matter densities from spherical Jeans fits.  
In particular, we modeled the rounder metal-poor component as it is less prone to modeling systematics due to flattening.
We derived $\log_{10}{(J(0.5\degree)/{\rm Gev^{-2}\, cm^{-5}})}\approx19.1$ and $\log_{10}{(J(0.5\degree)/{\rm Gev^{-2}\, cm^{-5}})}\approx 18.6$   with the Keck and MMT data set, respectively.
Thus, UMi is an excellent target for searches for dark matter annihilation or decay.

The presence of a highly flattened metal-rich population in UMi is unexpected and deserves a closer look. If the hints for prolate rotation are confirmed, that will add to the puzzle of the formation and evolution of UMi. The methods we have developed for the analysis of UMi can be applied to other classical dSphs to characterize their multiple populations and constrain their dark matter profiles more robustly.

\section*{Acknowledgments}

We thank Matt Walker for sharing an earlier catalog of the UMi MMT/Hectochelle data. 
We thank James Bullock, Mike Cooper, Sergey Koposov, Jen Marshall, Louie Strigari, and Matt Walker for helpful comments and discussion.
We thank the referee for their helpful comments.

ABP was supported by a GAANN fellowship at UCI.
ABP acknowledges generous support from the George P. and Cynthia Woods Institute for Fundamental Physics and Astronomy at Texas A\&M University.
ABP is supported by NSF grant AST-1813881.
ENK gratefully acknowledges support from a Cottrell Scholar award administered by the Research Corporation for Science Advancement as well as funding from generous donors to the California Institute of Technology.
RRM acknowledges partial support from project BASAL AFB-$170002$ as well as FONDECYT project N$^{\circ}1170364$. 
SGD was supported in part by the NSF grants AST-1413600, AST-1518308, and AST-1749235.

The authors wish to recognize and acknowledge the very significant cultural role and reverence that the summit of Mauna Kea has always had within the indigenous Hawaiian community.  We are most fortunate to have the opportunity to conduct observations from this mountain.

Funding for SDSS-III has been provided by the Alfred P. Sloan Foundation, the Participating Institutions, the National Science Foundation, and the U.S. Department of Energy Office of Science. The SDSS-III web site is \url{http://www.sdss3.org/}.

SDSS-III is managed by the Astrophysical Research Consortium for the Participating Institutions of the SDSS-III Collaboration including the University of Arizona, the Brazilian Participation Group, Brookhaven National Laboratory, Carnegie Mellon University, University of Florida, the French Participation Group, the German Participation Group, Harvard University, the Instituto de Astrofisica de Canarias, the Michigan State/Notre Dame/JINA Participation Group, Johns Hopkins University, Lawrence Berkeley National Laboratory, Max Planck Institute for Astrophysics, Max Planck Institute for Extraterrestrial Physics, New Mexico State University, New York University, Ohio State University, Pennsylvania State University, University of Portsmouth, Princeton University, the Spanish Participation Group, University of Tokyo, University of Utah, Vanderbilt University, University of Virginia, University of Washington, and Yale University.

This work has made use of data from the European Space Agency (ESA)
mission {\it Gaia} (\url{https://www.cosmos.esa.int/gaia}), processed by the {\it Gaia} Data Processing and Analysis Consortium (DPAC,
\url{https://www.cosmos.esa.int/web/gaia/dpac/consortium}). Funding
for the DPAC has been provided by national institutions, in particular
the institutions participating in the {\it Gaia} Multilateral Agreement.

This research has made use of NASA's Astrophysics Data System Bibliographic Services.

Databases and software: Besan\c{c}on model\footnote{\url{http://model.obs-besancon.fr/}} \citep{Robin2003}.
Python packages: \texttt{Astropy}\footnote{\url{http://www.astropy.org}} \citep{astropy2013},  \texttt{NumPy} \citep{numpy}, \texttt{iPython} \citep{ipython}, \texttt{SciPy} \citep{scipy}, and \texttt{matplotlib} \citep{matplotlib}, \texttt{corner.py} \citep{corner},
\texttt{emcee} \citep{ForemanMackey2013PASP..125..306F}

\bibliographystyle{mnras}
\bibliography{main_bib_file}




\bsp	
\label{lastpage}
\end{document}

%% file: table_data.tex
\begin{table*}
\ra{1.3}
\caption{List of Keck/DEIMOS  velocity and metallicity measurements.
Columns: (1) ID (2) {\it Gaia} DR2 source\_id. Stars with -1 are not in the {\it Gaia} catalog. (3) Megacam ID \citep{Munoz2018ApJ...860...65M}. Stars with -1 are not in the Megacam catalog. (4) RA (deg) (J2000) (5) DEC (deg) (J2000) (6) Modified JD (MJD) (7) $v_{\rm los}\,(\kms)$ (8) [Fe/H] (9) dSph membership (10) Metal-rich population membership (111) comments.  CMD = excluded due to location on color-magnitude diagram. Gaia NM = Non-member due to non-zero parallax and/or large proper motion. RRL = RR Lyrae star in {\rm Gaia} or PS1 catalog. NA8190 = MW foreground star due to Na {\rm \scriptsize I} doublet. This table is available in its entirety in the electronic edition of the journal. A portion is
reproduced here to provide guidance on form and content.
}
\begin{center}
\begin{tabular}{l ccc ccc ccc ccc}
\toprule
ID & {\it Gaia} DR2 source\_id & megacam\_id & RA (deg) & DEC (deg) & MJD & $v_{\rm los}$ ($\kms$) & [Fe/H] & $p_{\rm dSph}$ & $p_{\rm MR}$ &  comments \\ 
\midrule
1 & 1645443305863662592 & 33510 & 227.542184 & 67.177103 & 54884.5 & -222.6 $\pm$ 2.3 & -2.57 $\pm$ 0.11 & 1.00 & 0.00 & \\
   2 & 1645448979516115712 & -1 & 227.494207 & 67.272232 & 54884.5 & -242.7 $\pm$ 2.2 & -1.55 $\pm$ 0.10 & 1.00 & 0.90 & \\
   3 & 1645447811285006464 & -1 & 227.537617 & 67.214524 & 54884.5 & -233.4 $\pm$ 2.2 & -1.74 $\pm$ 0.10 & 1.00 & 0.44 & \\
   4 & 1645485263399853696 & -1 & 227.612946 & 67.410042 & 54884.5 & -250.1 $\pm$ 2.2 & -2.07 $\pm$ 0.10 & 1.00 & 0.75 & \\
   5 & 1645449426192720768 & 32845 & 227.580337 & 67.304530 & 54884.5 & -7.7 $\pm$ 2.2 & -1.88 $\pm$ 0.10 &  -  &  -  & CMD; Gaia NM; NA8190\\
\bottomrule
\end{tabular}
\end{center}
\par
\label{table:data}
\end{table*}

%% file: table_properties.tex
\begin{table*}
\ra{1.3}
\caption{ Posterior distributions for one and two population models for the RGB sample.  
The 1 and 2 labels refer to the metal-rich and metal-poor populations, respectively.
The second and third column are results with the Keck/DEIMOS data set and the last two columns are results with  the MMT/Hechocelle data set \citep{Spencer2018AJ....156..257S}.  
Note: $R_h = r_p \sqrt{1 - \epsilon}$.  
}
\begin{center}
\begin{tabular}{l r r r r}
\toprule
Parameter & Single (Keck) & Two (Keck) & Single (MMT) & Two (MMT) \\
\midrule
\hline
$\overline{v} \, ({\rm km \, s^{1}})$ & $-244.7_{-0.4}^{+0.4}$ & $-244.7_{-0.3}^{+0.3}$ & $-247.0_{-0.4}^{+0.4}$ & $-246.9_{-0.4}^{+0.4}$ \\ 
$\mu_{\alpha *}$ & $-0.149_{-0.014}^{+0.014}$ & $-0.149_{-0.014}^{+0.014}$ & $-0.150_{-0.012}^{+0.012}$ & $-0.150_{-0.012}^{+0.012}$ \\ 
$\mu_{\delta}$ & $0.064_{-0.013}^{+0.013}$ & $0.064_{-0.013}^{+0.013}$ & $0.053_{-0.011}^{+0.011}$ & $0.053_{-0.011}^{+0.011}$ \\ 
$\sigma_{v, 1} (\kms)$ &$8.6_{-0.3}^{+0.3}$ & $4.9_{-1.0}^{+0.8}$ & $8.6_{-0.3}^{+0.3}$ & $7.3_{-0.6}^{+0.5}$ \\ 
$\sigma_{v, 2} (\kms)$ & -- & $11.5_{-0.8}^{+0.9}$ &  -- & $11.7_{-1.0}^{+1.2}$ \\ 
$\overline{[{\rm Fe/H}]}_{1}$ &$-2.15_{-0.02}^{+0.01}$ & $-2.05_{-0.03}^{+0.03}$ & $-2.02_{-0.02}^{+0.02}$ & $-1.90_{-0.03}^{+0.03}$ \\ 
$\overline{[{\rm Fe/H}]}_{2}$ & -- & $-2.29_{-0.06}^{+0.05}$ &  -- & $-2.41_{-0.11}^{+0.10}$ \\ 
$\sigma_{\rm [Fe/H], 1}$ & $0.34_{-0.01}^{+0.01}$ & $0.26_{-0.03}^{+0.02}$ & $0.37_{-0.02}^{+0.02}$ & $0.26_{-0.02}^{+0.02}$ \\ 
$\sigma_{\rm [Fe/H], 2}$ & -- & $0.36_{-0.03}^{+0.03}$ &  -- & $0.35_{-0.04}^{+0.05}$ \\ 
$r_{p, 1} \, ({\rm pc})$ &$449_{-24}^{+27}$ & $444_{-37}^{+42}$ & $450_{-30}^{+33}$ & $425_{-38}^{+45}$ \\ 
$r_{p, 2} \, ({\rm pc})$ & -- & $457_{-46}^{+58}$ &  -- & $582_{-121}^{+185}$ \\ 
$\epsilon_{ 1} $ &$0.60_{-0.03}^{+0.02}$ & $0.75_{-0.03}^{+0.03}$ & $0.56_{-0.03}^{+0.03}$ & $0.64_{-0.04}^{+0.04}$ \\ 
$\epsilon_{2} $ & -- & $0.33_{-0.12}^{+0.09}$ &  -- & $0.21_{-0.14}^{+0.16}$ \\ 
$R_{h, 1} \, ({\rm pc})$ &$286_{-12}^{+13}$ & $221_{-17}^{+17}$ & $297_{-15}^{+15}$ & $253_{-17}^{+18}$ \\ 
$R_{h, 2} \, ({\rm pc})$ & -- & $374_{-37}^{+49}$ &  -- & $512_{-97}^{+145}$ \\ 
$f_{1}$ & -- & $0.54_{-0.10}^{+0.09}$ &  -- & $0.76_{-0.09}^{+0.07}$ \\
\bottomrule
\end{tabular}
\end{center}
\par
\label{table:results}
\end{table*}

%% file: table_mmt.tex
\begin{table}
\ra{1.3}
\caption{UMi membership of MMT/Hectochelle sample \citep{Spencer2018AJ....156..257S}.  
Columns: (1) {\it Gaia} DR2 source\_id, (2) membership in dSph MW model ($p_{\rm dSph}$) (3) membership in metal-rich population ($p_{\rm MR}$) (4) Comments.  CMD = star outside of color-magnitude selection box and excluded from analysis. Gaia = non-zero parallax and/or large proper motion and considered MW star. This table is available in its entirety in the electronic edition of the journal. A portion is
reproduced here to provide guidance on form and content.
}
\begin{center}
\begin{tabular}{l ccc }
\toprule
{\it Gaia} DR2 source\_id & $p_{\rm dSph}$ & $p_{\rm MR}$ & comments \\ 
\midrule
\hline
1645329064029028992 & 0.00 &  -  & CMD; Gaia\\
1645332259484700800 & 1.00 & 0.13 & \\
1645337580948516736 & 0.00 &  -  & Gaia\\
1645337958905643648 & 1.00 & 0.19 & \\
1645338130704328064 & 1.00 & 0.61 & \\
1645339024057541120 & 1.00 & 0.77 & \\
\bottomrule
\end{tabular}
\end{center}
\par
\label{table:mmt}
\end{table}

%% file: table_jfac.tex
\begin{table*}
\ra{1.3}
\caption{The J- and D-factors for UMi integrated over several solid angles with the RGB sample and the metal-poor sample for the Keck/DEIMOS and MMT/Hectochelle data sets. The J- and D-factors have units of $\mathrm{GeV}^2 \, \mathrm{cm}^{-5}$ and $\mathrm{GeV} \, \mathrm{cm}^{-2}$, respectively.  For reference, $1\, M_{\sun}^2 \mathrm{kpc}^{-5} = 4.45 \times 10^{6} \, \mathrm{GeV}^2 \, \mathrm{cm}^{-5}$ and $1\, M_{\sun} \mathrm{kpc}^{-2} = 1.17 \times 10^{14} \, \mathrm{GeV} \, \mathrm{cm}^{-2}$. We list the $1-\sigma$ and $2-\sigma$ error bars.
}
\begin{center}
\begin{tabular}{l cc ccc cc}
\toprule
Model &  $\log_{10}{J(0.1\degree))}$ &
$\log_{10}{J(0.2\degree))}$ & $\log_{10}{J(0.5\degree))}$ &
$\log_{10}{J(1.0\degree))}$ &
$\log_{10}{D(0.1\degree))}$ & $\log_{10}{D(0.2\degree))}$ &
$\log_{10}{D(0.5\degree))}$ \\
\midrule
Keck - MP & $18.80_{-0.33(0.68)}^{+0.30(0.58)}$ & $18.98_{-0.21(0.43)}^{+0.21(0.45)}$ & $19.08_{-0.13(0.23)}^{+0.16(0.37)}$ & $19.11_{-0.12(0.23)}^{+0.16(0.36)}$ & $17.51_{-0.06(0.13)}^{+0.05(0.10)}$ & $17.93_{-0.05(0.09)}^{+0.06(0.14)}$ & $17.93_{-0.18(0.32)}^{+0.19(0.37)}$\\
Keck - RGB & $18.52_{-0.41(0.72)}^{+0.31(0.51)}$ & $18.64_{-0.22(0.38)}^{+0.24(0.42)}$ & $18.73_{-0.11(0.18)}^{+0.17(0.34)}$ & $18.78_{-0.12(0.20)}^{+0.17(0.33)}$ & $17.31_{-0.03(0.07)}^{+0.03(0.08)}$ & $17.70_{-0.05(0.09)}^{+0.11(0.22)}$ & $17.64_{-0.23(0.34)}^{+0.28(0.53)}$\\
MMT - MP & $18.01_{-0.59(1.15)}^{+0.55(1.04)}$ & $18.34_{-0.46(0.94)}^{+0.43(0.83)}$ & $18.61_{-0.31(0.62)}^{+0.31(0.63)}$ & $18.71_{-0.25(0.48)}^{+0.26(0.56)}$ & $17.27_{-0.16(0.32)}^{+0.16(0.30)}$ & $17.77_{-0.13(0.26)}^{+0.12(0.24)}$ & $17.94_{-0.20(0.47)}^{+0.14(0.27)}$\\
MMT - RGB & $18.16_{-0.22(0.48)}^{+0.28(0.59)}$ & $18.37_{-0.14(0.32)}^{+0.18(0.42)}$ & $18.47_{-0.09(0.18)}^{+0.13(0.35)}$ & $18.49_{-0.09(0.17)}^{+0.13(0.34)}$ & $17.21_{-0.05(0.11)}^{+0.05(0.09)}$ & $17.64_{-0.03(0.06)}^{+0.03(0.07)}$ & $17.70_{-0.11(0.22)}^{+0.12(0.27)}$\\
\bottomrule
\end{tabular}
\end{center}
\par
\label{table:j_fac}
\end{table*}